\documentclass[%
 aip,
 jmp,%
 amsmath,amssymb,
 preprint,%
prl]{revtex4-1}

\usepackage{graphicx}
\usepackage{dcolumn}
\usepackage{bm}
\begin{document}

\title[Statistical properties of CHM zonal flows]
      {Statistical properties of Charney-Hasegawa-Mima zonal flows} 

\author{Johan Anderson}%
\email{anderson.johan@gmail.com.}
\affiliation{ 
Department of Earth and Space Sciences, Chalmers University of Technology, 
SE-412 96 G\"{o}teborg, Sweden
}%

\author{G. J. J. Botha}
\affiliation{ 
Department of Mathematics and Information Sciences, 
Northumbria University, Newcastle upon Tyne, NE1 8ST, 
United Kingdom 
}%

\date{\today}

\begin{abstract}
A theoretical interpretation of numerically generated probability
density functions (PDFs) of intermittent plasma transport events in unforced 
zonal flows is provided within the Charney-Hasegawa-Mima (CHM) model. 
The governing equation is solved numerically with various prescribed 
density gradients that are designed to produce different configurations 
of parallel and anti-parallel streams. Long-lasting vortices form whose flow 
is governed by the zonal streams. It is found that the numerically generated PDFs 
can be matched with analytical predictions of PDFs based on the instanton 
method by removing the autocorrelations from the time series. In many 
instances the statistics generated by the CHM dynamics relaxes to Gaussian 
distributions for both the electrostatic and vorticity perturbations, whereas 
in areas with strong nonlinear interactions it is found that the PDFs 
are exponentially distributed.
\end{abstract}

\maketitle


\section{Introduction}
\label{sec:intro}

In recent experiments and numerical simulations it has been found that
significant transport might be mediated by coherent structures such as
streamers, blobs and vortices through the formation of rare avalanche-like
events of large amplitude \cite{zweben2007, politzer2000, beyer2000, drake1988,
antar2001, carreras1996, nagashima2011, dif2010}. These events cause the
probability distribution function (PDF) to deviate from a Gaussian profile on
which the traditional mean field theory such as transport coefficients is based.
Specifically, the PDF tails manifest the intermittent character of transport due
to rare events of large amplitude that are often found to substantially differ
from Gaussian distribution, although PDF centres tend to be Gaussian. Therefore,
a comprehensive predictive theory is called for in order to understand and
subsequently improve intermittent transport features e.g. confinement degradation in
tokamaks. 

Drift wave turbulence is known to generate zonal flows, which in turn 
inhibits the growth of turbulence and transport 
\cite{DiamondEA2005,DiamondEA2011}. 
As such, zonal flows play an important role in fusion plasmas 
\cite{ConnorEA2004,ItohEA2006,ConnorMartin2007}. 
 
In geophysical fluid dynamics, zonal flows are believed to cause a similar 
reduction in transport in atmospheres \cite{SmedmanEA2004,PrabhaEA2008} 
under certain limiting conditions \cite{DuarteEA2012}. 
This comes as no surprise, given the analogy between drift waves in the 
dissipationless limit and Rossby waves in nearly incompressible, shallow 
rotating fluids; both systems are described by the 
Charney-Hasegawa-Mima (CHM) equation \cite{hi1996}.

Numerical studies where sheared flow is externally prescribed  
\cite{CarrerasEA1992,GarbetEA2002,TynanEA2006} 
lead to energetics that are qualitatively different from those 
obtained in the drift-wave/zonal-flow feedback mechanism 
\cite{ConnaughtonEA2011}. 
In the CHM model \cite{charney1948, hm1977,hm1978}
sheared flow may be imposed by prescribing the background density gradient
\cite{botha1999}. The known solutions of the CHM equation, namely 
bipolar and monopolar vortices \cite{Nycander1988}, form in the plasma.
Here the term vortex is used to describe a localized extremum in the 
electrostatic potential that is evolved through the CHM equation. 
In this paper fluid simulations using the CHM equation are designed so 
that zonal flows 

are prescribed externally, 
 
moving in the poloidal direction and with various 
configurations in the radial direction. 
At initialisation bipolar vortices form, but only monopolar vortices 
survive due to the interaction between vortices as well as the destructive 
effects of the sheared flow \cite{hi1996}. 

The fluid simulations produce quasi-stationary time series (poloidally averaged and sampled at different radial points) of the electrostatic potential and corresponding vorticity that describe the CHM flows \cite{horton1999, charney1948, hm1977, hm1978, hmk1979, botha1999}. We apply a standard Box-Jenkins modeling for each time series. This mathematical procedure effectively removes deterministic autocorrelations from the time series, allowing for the statistical interpretation of the stochastic residual part. In this particular case it turns out that an ARIMA(3,1,0) model (autoregressive integrated moving average) \cite{box1994} accurately describes the stochastic process.

The stochastic residual of the time series of potential and vorticity exhibit Gaussian statistics or distributions with elevated exponential tails. We utilize analytical results from nonperturbative stochastic theory, the so-called instanton method \cite{justin1989, gurarie1996, falcovich1996, kim2002, anderson1, kim2008, anderson2} for computing PDFs in turbulence as a comparison to the numerical data. The analytically derived PDFs are rather insensitive to the details of the linear physics of the system \cite{kim2008} and thus display salient features of the nonlinear interactions. The numerically generated time traces are analysed using the ARIMA model and fitted with the analytical models accordingly. We find in the regions with strong nonlinear characteristics an emergent universal scaling of the PDF tails of exponential form $\sim\exp\big(- const \ |\phi|\big)$ as suggested by recent theoretical work in Ref. \onlinecite{falcovich2011, kim2008, anderson4} relevant for the direct cascade dynamics. However, in many cases for the CHM zonal flows we find Gaussian PDFs in similarity to what was seen from the theoretical model in Ref. \onlinecite{anderson3}, whereas for some mid radial points the system exhibit sub-exponential PDFs $\sim\exp\big(-const \ |\phi|^{\alpha}\big)$ with $\alpha < 1$, where the dynamics is strongly influenced by the zonal flow resulting in strong intermittency.

The paper is organized as follows: Section \ref{sec:CHM} introduces the 
CHM model and Section \ref{sec:level3} the statistical method used in the analysis. The numerical results are presented in Section \ref{sec:results} after which they are discussed and the paper is concluded by a summary of the main results. 


\section{Charney-Hasegawa-Mima model}
\label{sec:CHM}

The CHM equation is solved in the Cartesian plane perpendicular 
to a constant magnetic field ${\bf B}=B_0{\bf \hat{z}}$, 
with $x$ and $y$ being the radial and poloidal coordinates of toroidal 
geometry respectively. 
The plane is periodic in the $y$ direction and finite 
in the $x$ direction.
The background electrostatic field $\phi$ evolves through time 
on the ($x,y$) plane through the CHM equation 
\cite{charney1948,hm1977,hm1978,hmk1979}
\begin{equation}
\frac{L_n}{c_s}\frac{\partial}{\partial t}
\left(\phi-\rho_s^2\nabla_\perp^2\phi\right)
-\rho_s\frac{\partial\phi}{\partial y}
-\frac{\rho_s}{v_*}\frac{c}{B}
\left[\phi,\rho_s^2\nabla_\perp^2\phi\right] = 0, 
\label{eq:CHM}
\end{equation}
where $[\;,\:]$ are the Poission brackets. The 
diamagnetic velocity ${\bf v}_\star=v_\star{\bf\hat{y}}$ 
is defined as
\begin{equation}
v_* =
\frac{cT_e}{eB}\frac{1}{n_0}\frac{\partial n_0}{\partial x}
         =\frac{c_s^2}{\Omega_{ci}}L_n^{-1} ,
\label{eq:vstar}
\end{equation}
with $c$ the speed of light, $T_e$ the electron temperature, 
$e$ the electron charge, $c_s$ the sound speed, $\Omega_{ci}$ 
the ion cyclotron frequency and $n_0$ the time-independent nonlinear 
background density. 
The characteristic length is the thermal Larmor radius 
$\rho_s=c_s/\Omega_{ci}$ and the characteristic time is taken as 
$\rho_s/\max|v_\star|=(c_s\max|L_n^{-1}|)^{-1}$. 
Equation (\ref{eq:CHM}) has two global invariants: the generalised energy $W$ 
and the generalised enstrophy $U$. 

A second-order modified Euler predictor-corrector time scheme is used to 
solve equation (\ref{eq:CHM}). The periodic $y$ direction is treated 
spectrally while the $x$ direction as well as the nonlinearity of 
equation (\ref{eq:CHM}) are finite differenced \cite{botha1999}. The 
CHM equation's nonlinear term is calculated using a conservative scheme 
for vector nonlinearities \cite{arakawa1966}. The $x$ boundary 
conditions are fixed with $\phi=0$ at all time. 

The numerical runs are initialized with a perturbation along the 
$y$ direction consisting of many wavelengths, while the nonlinear density gradient $L_n^{-1}=L_n^{-1}(x)$ is prescribed and kept constant. The 
CHM equation produces a solution containing many pairs of bipolar vortices
that evolve into larger monopolar vortices, the latter existing for 
most of the numerical run. In order to do a statistical analysis on these 
fluctuations, time series of the algebraic averages in the poloidal 
($y$) direction of the normalised electrostatic potential $e\phi/T_e$
and normalised vorticity $\omega/\Omega_{ci}$ 
are obtained at positions along the radial ($x$) axis, denoted as 
$\bar{\phi}(x,t)$ and $\bar{\omega}(x,t)$.  
The normalised  vorticity is obtained using
\begin{equation}
\frac{\omega}{\Omega_{ci}} = 
\rho_s^2\nabla^2_{\perp}\frac{e\phi}{T_e}\; .
\end{equation}


\section{Statistical analysis}
\label{sec:level3}

In this paragraph we will quantify the intermittency in the simulated
time series by computing the PDFs of the residuals or the stochastic component
of the time traces and compare these with analytical predictions. Here, we
briefly outline the implementation of the instanton method. For more details,
the reader is referred to the existing literature \cite{justin1989}. In
the instanton method the PDF tail is first formally expressed in terms of a 
path integral by utilizing the Gaussian statistics of the forcing, in a 
similar spirit as in 
Refs. \onlinecite{justin1989, kim2002, anderson1, kim2008, anderson2}.
Here and throughout this paper, the term forcing is meant to describe the 
inherent unpredictability of the dynamics and will be assumed to be Gaussian for simplicity. 
A general class of solutions is presented in Ref. \onlinecite{kim2008}.
The integral in the action ($S_{\lambda}$) in the path integral is evaluated using the saddle-point 
method in the limit $\lambda \rightarrow \infty$ representing the tail values.
The parameter $\lambda$ is proportional to some power of the quantity of interest 
such as the potential or flux. 
In mathematical terms, this corresponds to evaluating the integral along an 
optimum path among all possible paths or functional values. The instanton 
is localized in 
time, existing during the formation of coherent structure. The
saddle-point solution of the dynamical variable $\phi({\bf x},t)$ of the form
$\phi({\bf x},t) = F(t) \psi({\bf x})$ is called an instanton if $F(t) = 0$ at
$t=-\infty$ and $F(t) \neq 0$ at $t=0$. Note that, the function $\psi({\bf x})$
here represents the spatial form of the coherent structure. Thus, the
intermittent character of the transport consisting of bursty events can be
described by the creation of the coherent structures. The dynamical system with
a stochastic forcing is enforced to be satisfied by introducing a larger state
space involving a conjugate variable $\phi^*$, whereby $\phi$ and $\phi^*$ 
constitute an uncertainty relation. Furthermore, $\phi^*$ acts as a mediator 
between the observables (potential or vorticity) and instantons 
(physical variables) through
stochastic forcing. Based on the assumption that the total PDF can be
characterized by an exponential form and that it is symmetric around the mean
value $\mu$, the expression
\begin{eqnarray}
P(\phi) & = & \frac{1}{Nb} \exp{\{ - \frac{1}{b} |\phi-\mu |^{\chi}\}},
\label{pq2}
\end{eqnarray}
is found, where the potential $\phi$ plays the role of the stochastic variable,
with $P(\phi)$ determining its statistical properties. Here $b$ is a constant
containing the physical properties of the system. Using the instanton method we
find different statistics in different situations. In a vorticity conserving
system the intermittent properties of the time series in simulations are
attributed to rare events of modon like structures that have a simplified
response for the vorticity,
\begin{eqnarray} \label{modrel}
\nabla_{\perp}^2 \phi = - k_{\perp}^2 \phi + \eta x.
\end{eqnarray}

Here $\eta = 1 + (1 - k^2)U$ and the vortex speed is $U$. In this situation 
it has been predicted \cite{falcovich2011, anderson4} that the system  has exponential 
tails in the direct cascade, 
$\exp\big(- const \ |\omega|\big) \sim \exp\big(- const \ |\nabla_{\perp}^2
\phi|\big)  \sim \exp\big(-const \ |k_{\perp}^2 \phi|\big)$, indicating a 
value of $\chi=1.0$ as in Ref. \onlinecite{falcovich2011, anderson4}. 
In the References \onlinecite{kim2002, anderson1, anderson2} the 
 statistics of the momentum flux is found to be a stretched exponential with $\chi=3/2$. 
However, when the nonlinear interactions are weak, as well
as in the case of an imposed zonal flow we find Gaussian statistics where $\chi=2$ 
as is elucidated on in Ref. \onlinecite{anderson3}.
In the analysis we will make use of different types of distributions to retro-fit the
PDFs of simulation results mainly using the Laplace distribution ($\chi = 1.0$) and 
the Gaussian distribution ($\chi = 2.0$).

We focus on the time traces (averaged in the $y$-direction) at five 
equidistant radial points located at $x=-18.9, -9.5, 0.0, 9.5, 18.9$ 
(in units of $\rho_s$). 
Each set of data describes the time evolution of
the potential and vorticity to which we apply a standard Box-Jenkins
modeling~\cite{box1994}. This mathematical procedure effectively removes
deterministic autocorrelations from the system, allowing for the statistical
interpretation of the residual part, which a posteriori turns out to be relevant
for comparison with the analytical theory. In our set-up, it turns out that an
ARIMA(3,1,0) model accurately describes the stochastic procedure, in that, one
can express the (differenced) potential time trace in the form
\begin{eqnarray}
\phi_{t+1}=a_1\,\phi_{t}+a_2\,\phi_{t-1}+a_3\,\phi_{t-2}+\phi_{res}(t)
\end{eqnarray}
where the fitted coefficients $a_1, a_2, a_3$ describe the deterministic
component and $\phi_{res}$ is the residual part (noise or stochastic component).
In the time traces the mean values differ by several orders of magnitude 
and a convenient way of starting the comparison between different cases is
to apply rescaling of the data. In the rescaling procedure we multiply the 
original time trace with a constant factor, thus the mean and variance 
values are directly affected. However, the skewness and kurtosis are kept 
constant by construction. The benefits gained from rescaling are that we 
may compare a large number of different cases at different
radial points and that the tails are retained and the ARIMA model is preserved,
thus in this sense the original and rescaled data is statistically equivalent. 
The original simulation data sets are down-sampled and consists of typically
 $5\times 10^4$ entries. 


\section{Results}
\label{sec:results}

In this section the numerical results from all the different stream 
configurations are presented in tandem with the statistical analysis.
 
Throughout the paper the simulation plane has dimensions 
$-L_x\leq x\leq L_x$ and $0\leq y\leq L_y$ where 
$L_x=23.5\rho_s$ and $L_y=23.4\rho_s$. 
The characteristic length is $\rho_s=0.42$ cm.

\subsection{One stream}
\label{s:one}

\begin{figure}[t]
\centerline{
\includegraphics[width=8cm, height=5.5cm]{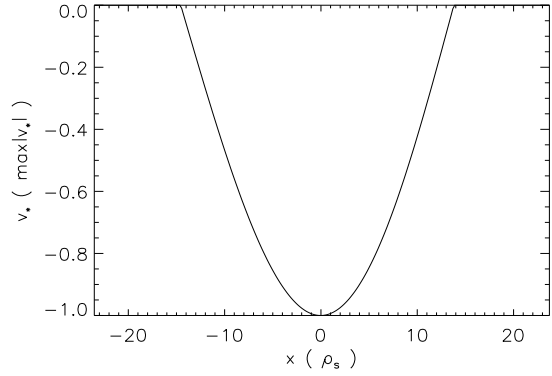}}
\caption{Constant $x$ profile of the normalised diamagnetic velocity 
$v_\star$. The characteristic time is 
$(c_sL_n^{-1})^{-1}=4.2\times 10^{-6}$ s. }
\label{f:one1}
\end{figure}

The constant background density gradient (Figure \ref{f:one1}) generates one 
stream centred at $x=0$ and with velocity $\max|v_\star|$ in the negative 
poloidal ($y$) direction. The flow is zero for $|x|\geq 15\rho_s$. The 
characteristic length and time in Figure \ref{f:one1} give 
$\max|v_\star|=10^5$ cm s$^{-1}$.

\begin{figure}[b]
\centerline{
\includegraphics[width=8.4cm, height=5.6cm]{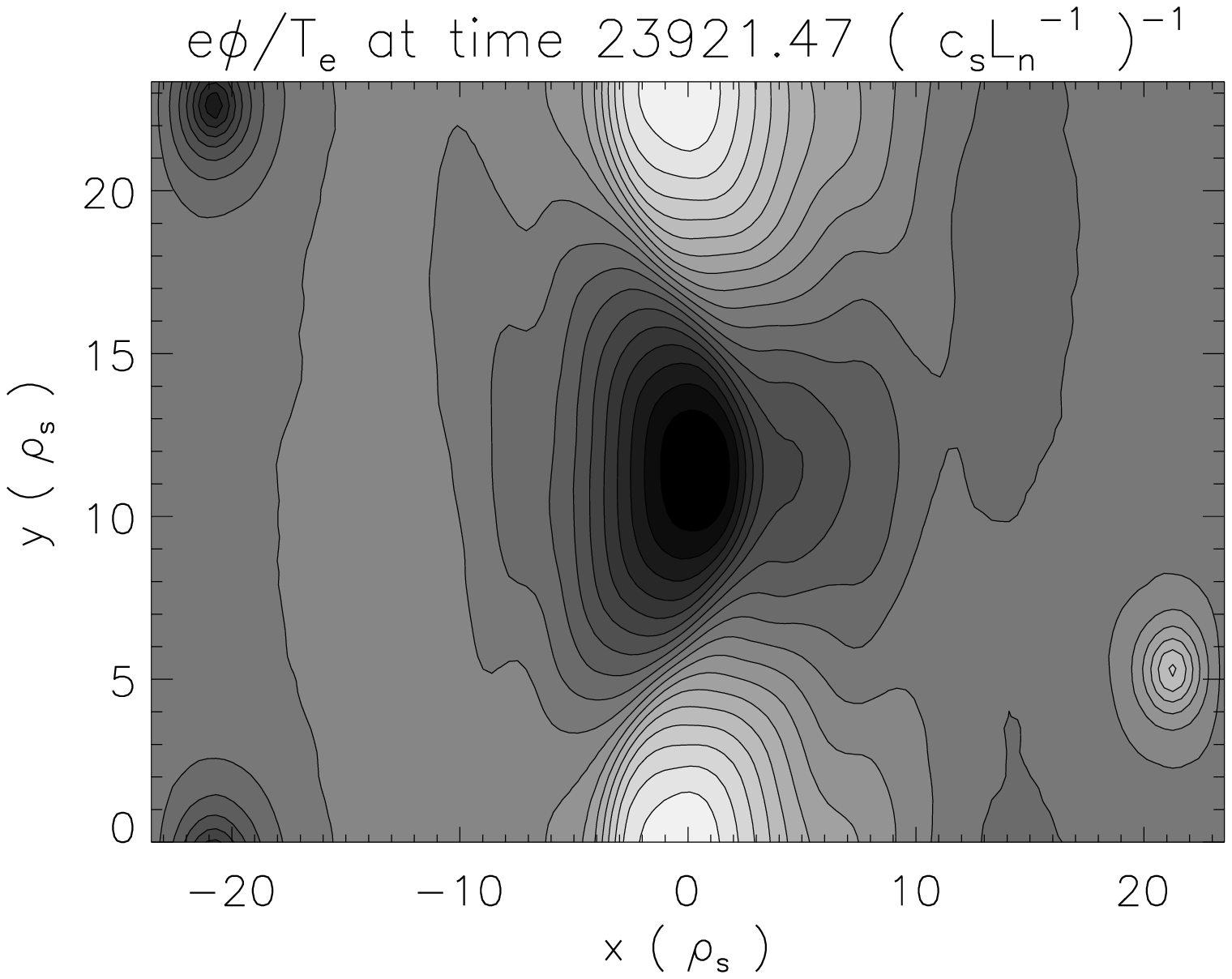}
\includegraphics[width=8.4cm, height=5.6cm]{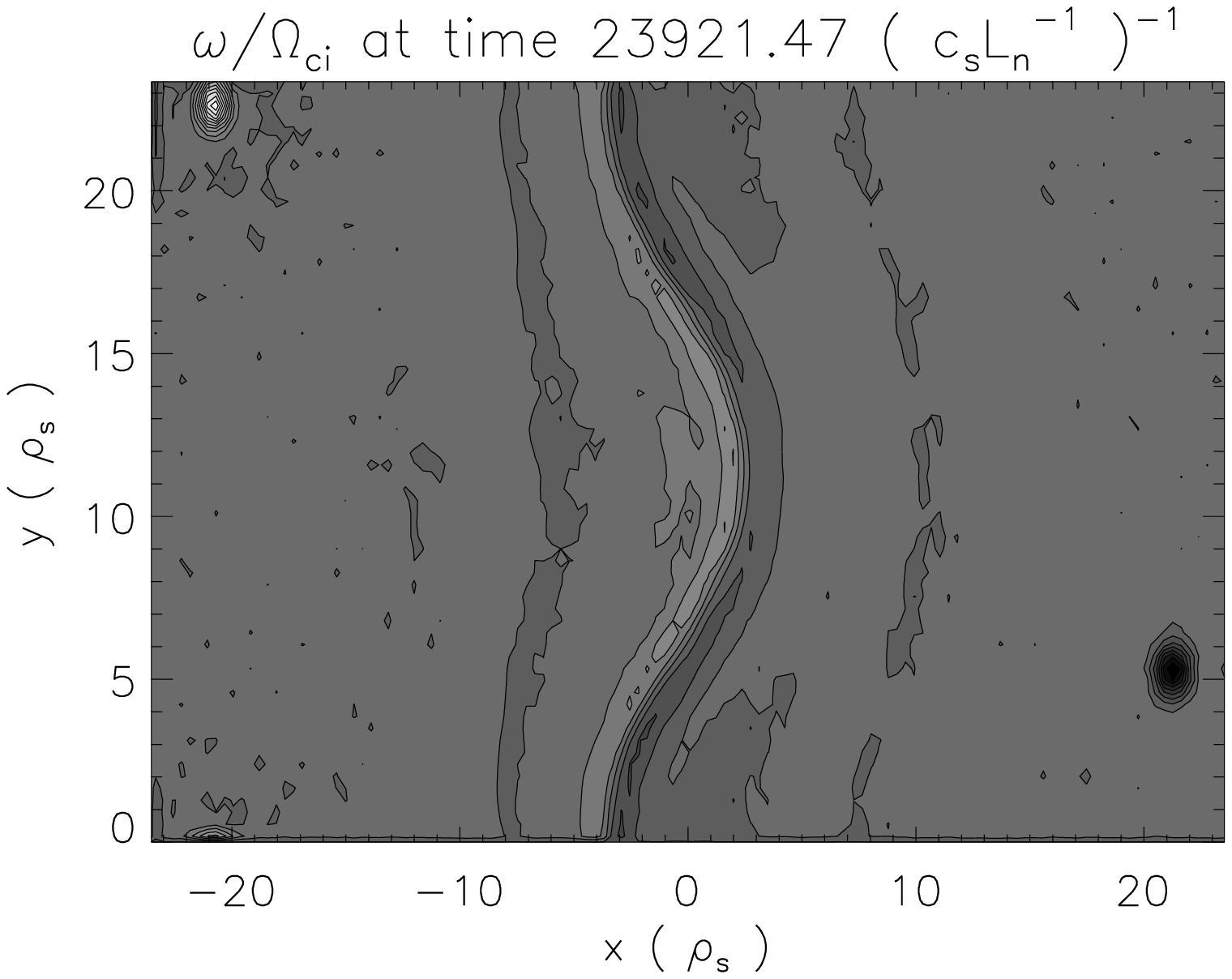} }
\caption{
The normalised potential $e\phi/T_e$ and normalised vorticity 
$\omega/\Omega_{ci}$ at 23921.47 characteristic time units for the one stream. 
The minimum and maximum values of $e\phi/T_e$ are $\pm0.019$ and the 
extrema for $\omega/\Omega_{ci}$ are -0.105 and 0.136.
Maximum is white and minimum black in both plots. The vortices
move in the negative $y$ direction at a speed of $0.2\max|v_\star|$.
}
\label{f:one2}
\end{figure}

At initialisation many vortices form that coalesce into large monopolar 
vortices, their widths determined by the width of the stream and 
with alternating polarities (Figure \ref{f:one2}). These monopolar vortices are 
dragged by the diamagnetic flow and move at a speed of $0.2\max|v_\star|$ in the 
negative $y$ direction, which is the direction of the diamagnetic velocity 
${\bf v}_\star$ (Figure \ref{f:one1}). 
Smaller vortices that earlier in the evolution moved outside the 
stream, move very slowly in random directions under the influence of the large 
vortices inside the stream, as can be seen at positions $x=\pm20\rho_s$.

\begin{figure}[t]
\centerline{
\includegraphics[width=8cm, height=5.5cm]{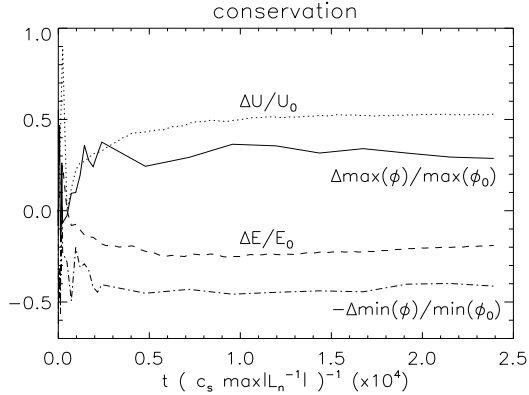}}
\caption{Conservation of the generalised energy (dashed line) and enstrophy
         (dotted line) for the one stream. The change in generalised energy 
         is $\bigtriangleup E/E_0 = (E_0-E_1)/E_0$ up to the end of the run 
         and the change in generalised enstrophy is 
         $\bigtriangleup U/U_0 = (U_0-U_1)/U_0$. 
         The notation $E_0$ and $U_0$ is used for the initial values and 
         $E_1$ and $U_1$ for the final values. 
         Table \ref{t:one} shows the conservation from times 
         $0.48\times 10^4$ 
         and $0.96\times 10^4$ $(c_s\max|L_n^{-1}|)^{-1}$ onwards.  
          }
\label{f:one3}
\end{figure}

After initialisation the generalised energy and enstrophy change 
significantly as vortices merge but once the large vortices have formed, 
from time $0.5\times 10^4$ $(c_s\max|L_n^{-1}|)^{-1}$ onwards,  
these quantities are relatively stable (Figure \ref{f:one3} and Table 
\ref{t:one}). Small changes in energy conservation are reflected in 
small changes in potential amplitudes. The change in enstrophy conservation 
is mirrored in changes in the vorticity of the fluctuations. 

\begin{table}
\begin{tabular}{lrrcrr}
\hline
&
\multicolumn{2}{c}{$\bigtriangleup E/E_t=(E_t-E_1)/E_t$} & $\quad$ &
\multicolumn{2}{c}{$\bigtriangleup U/U_t=(U_t-U_1)/U_t$}\\
& $t=0.2$ & $t=0.4$ & & $t=0.2$ & $t=0.4$ \\ \hline
One stream & $0.026$ & $0.171$ & & $-0.048$ & $-0.068$ \\
\hline
\end{tabular}
\vspace{1cm}
\caption
{Conservation of the generalised energy $E$ and generalised enstrophy $U$ 
for the one stream.
The time parameter $t$ is scaled such that $t=1$ is the end of the simulation.}
\label{t:one}
\end{table}

In the statistical analysis, the higher moments of the distribution function may reveal important features of the statistics and the underlying dynamics of the process. Here, in the statistical analysis we will in addition consider the kurtosis and skewness. We define the kurtosis as the fourth moment divided by the square of the second moment $kurtosis = m_4/m_2^2$, note that sometimes $3$ is subtracted from the kurtosis yielding a zero kurtosis for a Gaussian distribution. A high value of the kurtosis is a key mark for a heavy tailed distribution which is flat at the centre. The skewness is defined as the third moment normalized by the $3/2$-power of the second moment, $skewness = m_3/m_2^{3/2}$ and describes the asymmetry of the PDF around its mean value.  

\begin{figure}[ht]
{\vspace{2mm}
\includegraphics[width=5.7cm, height=5.6cm]{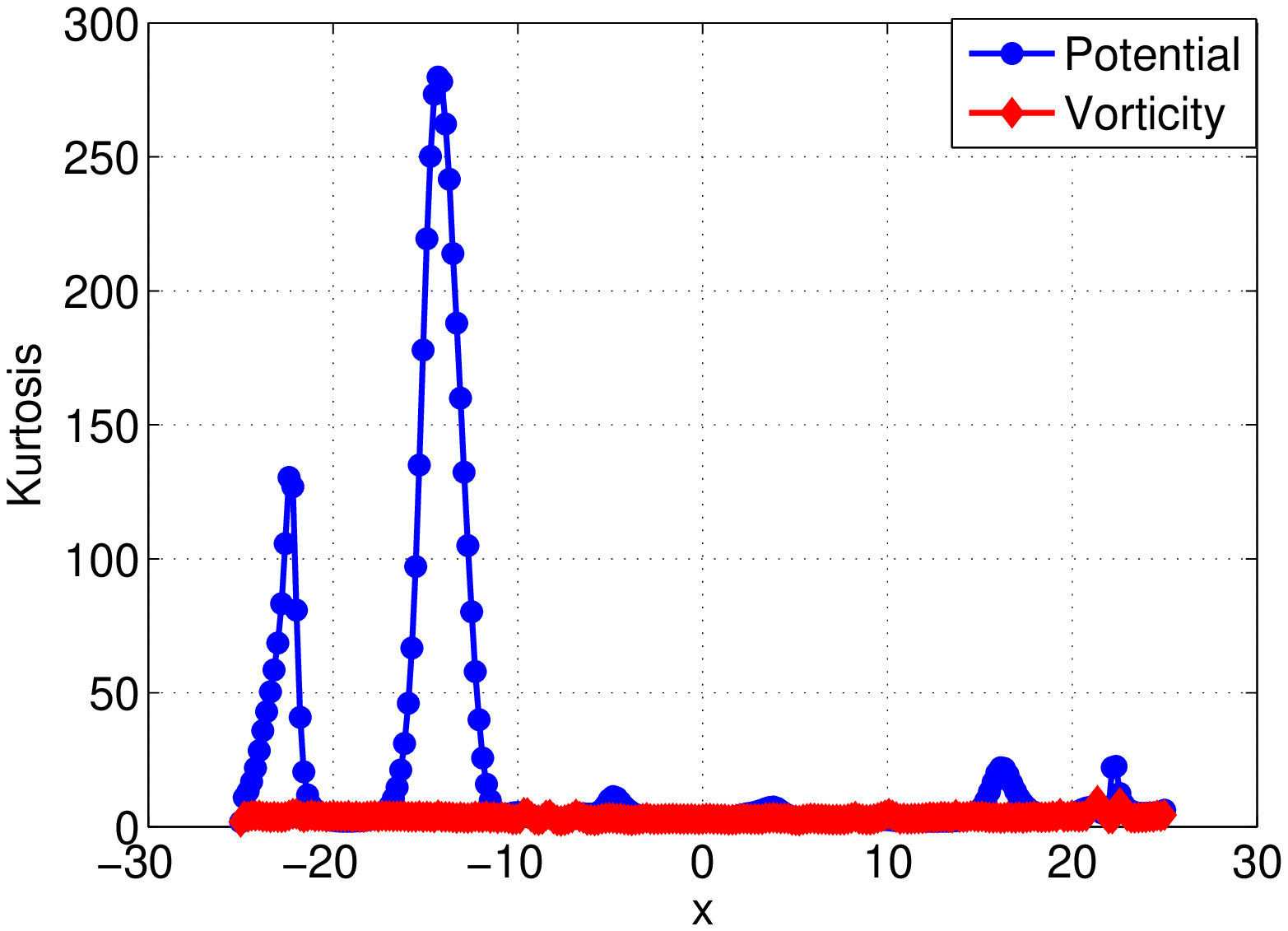}}
{\includegraphics[width=5.5cm, height=5.5cm]{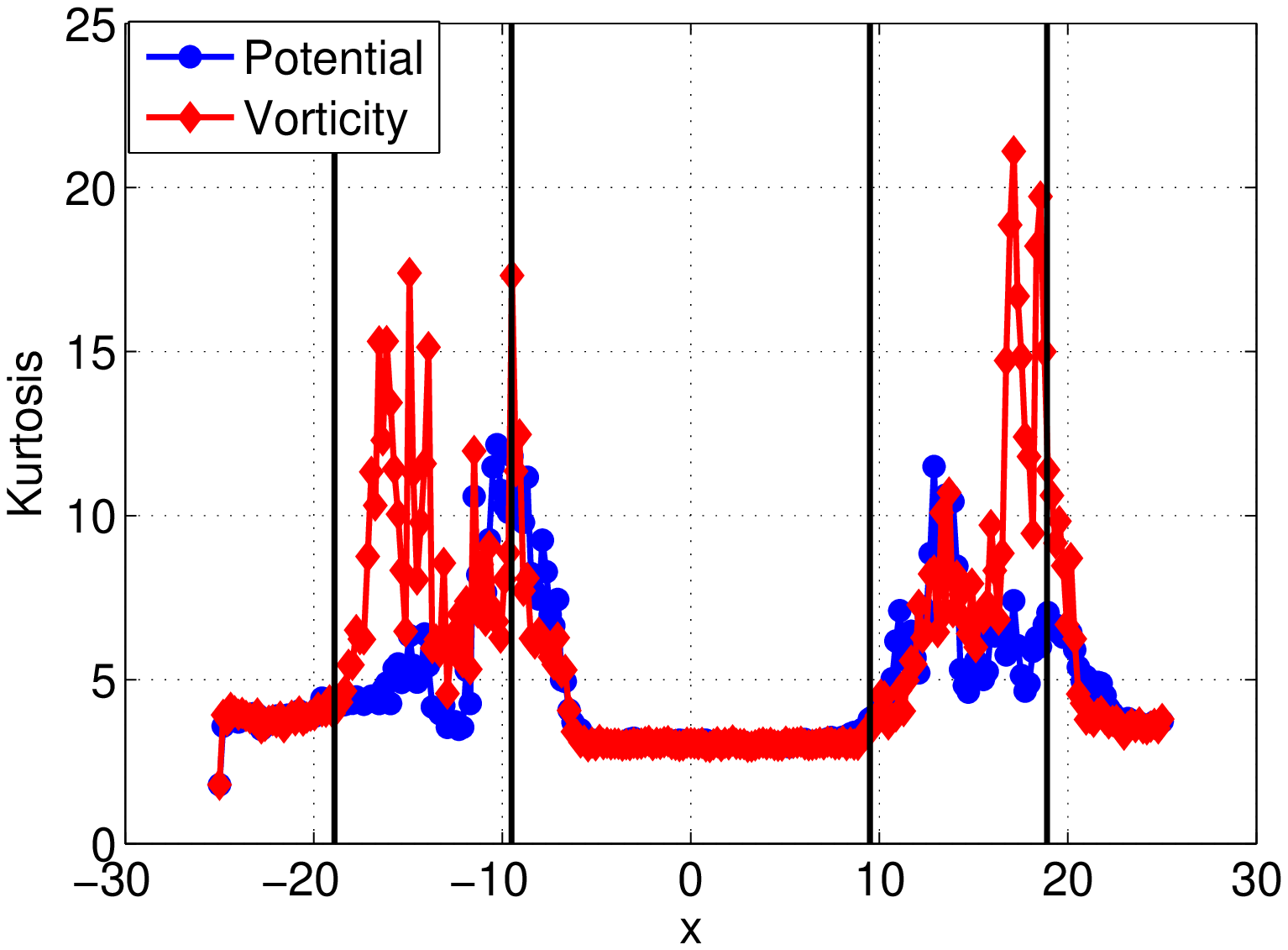}}
\caption{
The kurtosis of the potential and vorticity time traces along the x direction in the one stream case are shown for the original time traces (graph on the left) compared to the ARIMA modeled time traces (graph on the right).
}
\label{f:Kurt_one}
\end{figure}

In Figure \ref{f:Kurt_one}, the kurtosis along the $x$ direction for the original time series is compared to the ARIMA modeled residual stochastic part of the time series. At some negative $x$ locations distributions with elevated tails are found in the potential for the original time traces however this behaviour is not found for the vorticity. Furthermore, comparing the kurtosis of the potential and vorticity of the ARIMA modeled time traces the region with elevated tails coincide, as an indication of Eq. (\ref{modrel}).

\begin{figure}[ht]
{\vspace{2mm}
\includegraphics[width=8.1cm, height=8.5cm]{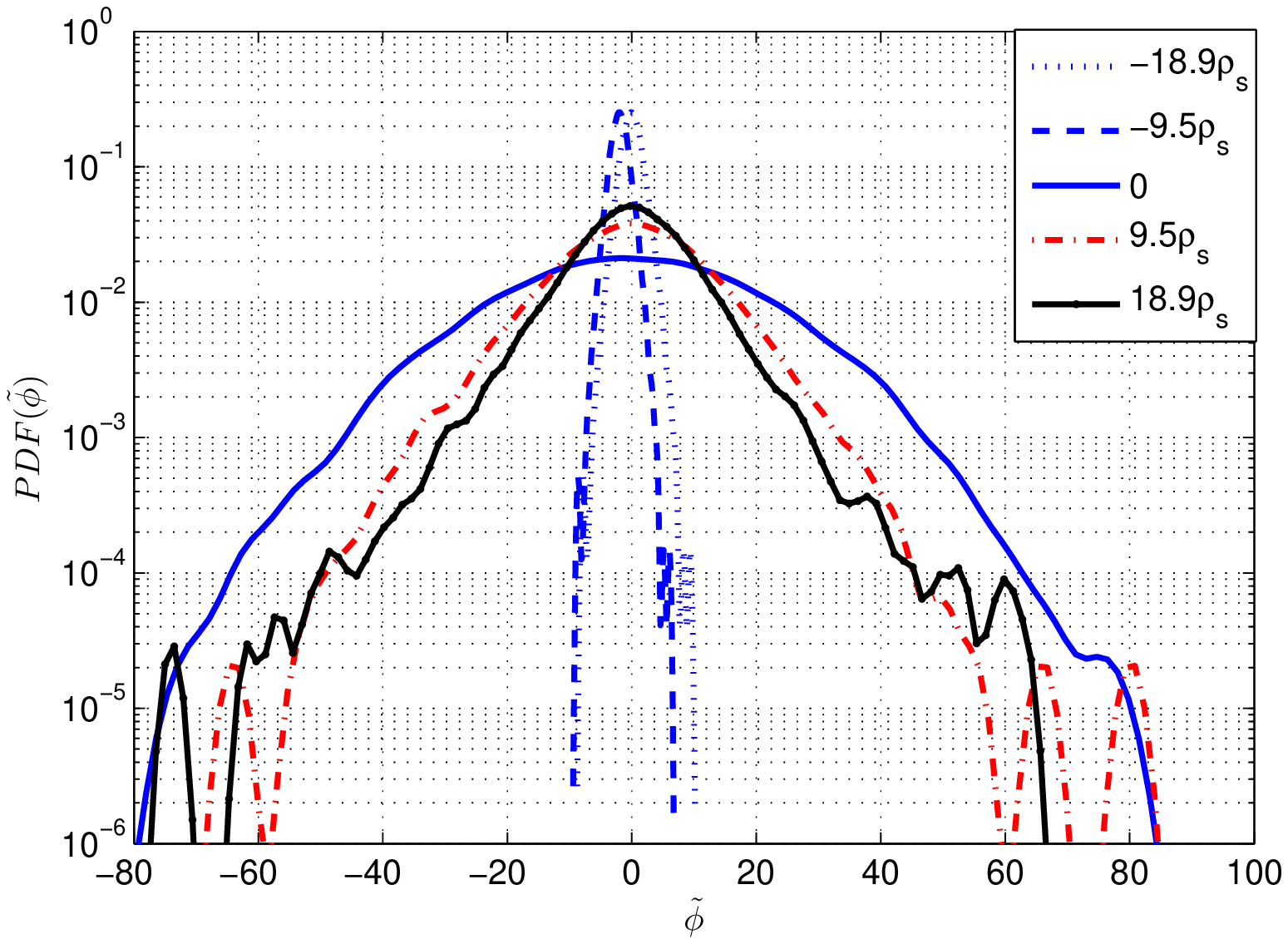}}
{\includegraphics[width=8.1cm, height=8.5cm]{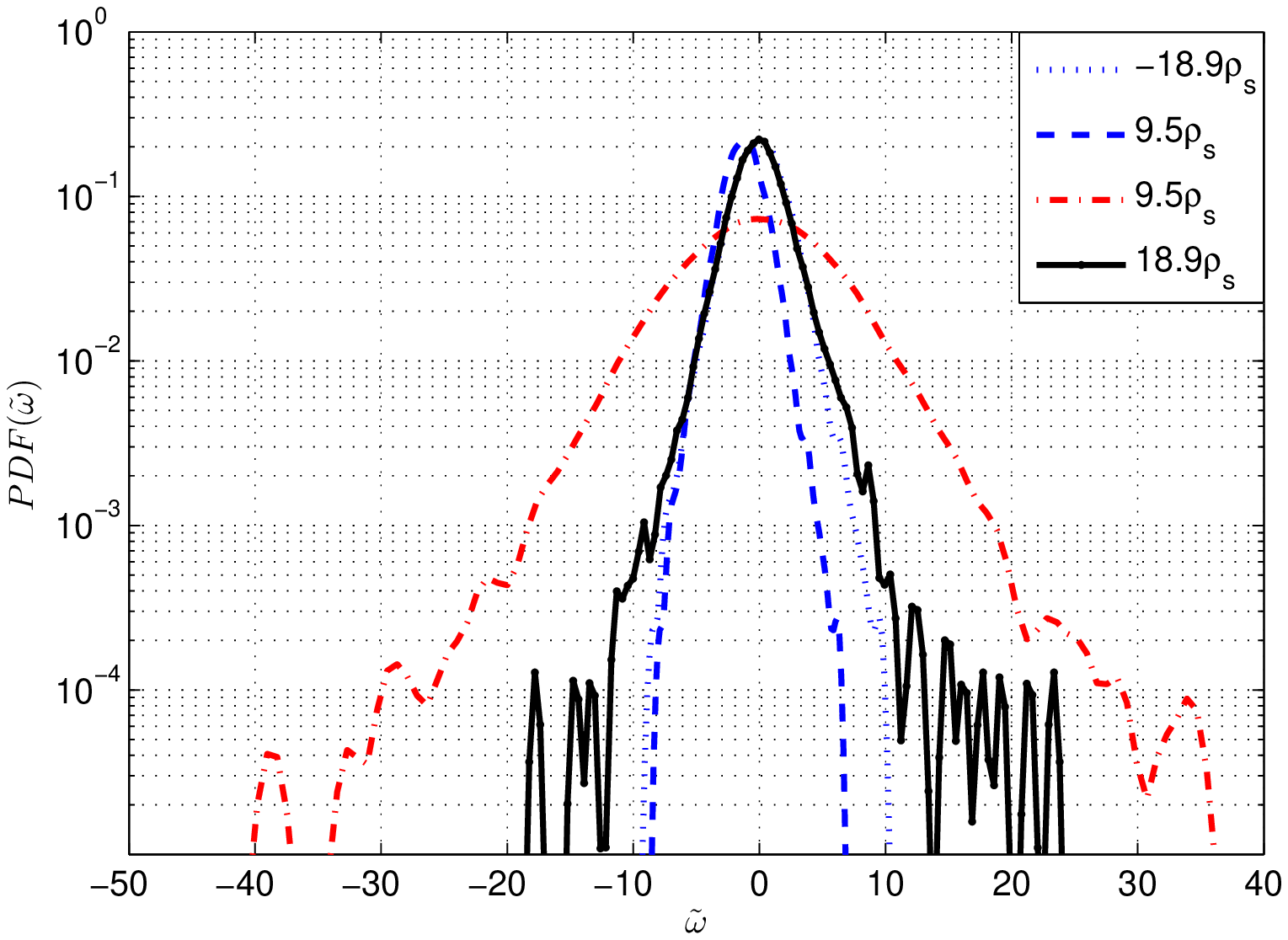}}
\caption{
Numerical PDFs along the x direction of the potential (graph on the left) and vorticity (graph on the right) time traces in the one stream case. The various x positions sampled are indicated by black vertical lines in Figure \ref{f:Kurt_one}. 
}
\label{f:PDF_one}
\end{figure}

Figure \ref{f:PDF_one} displays the PDFs at several positions marked by black lines in Figure \ref{f:Kurt_one} along the $x$ direction. The time traces are normalized in order to be able to show several PDFs in the same graph according to $PDF(\tilde{\phi}) = PDF[(\phi_{res} - \mu)/\mu]$, where $\mu$ is the mean value and $\phi_{res}$ is the residual after the ARIMA process. An analogous definition is adopted for the vorticity $\tilde{\omega}$. Here it can be seen that the PDFs of potential and kurtosis have exponential tails at the positions with higher kurtosis compared to the middle region which seems to exhibit Gaussian statistics. Note at some radial positions we find very large values of kurtosis signifying distributions with heavy tails and $\chi<1$. Moreover, the PDFs are nearly symmetric yielding small values of the skewness measure. In particular, Figure \ref{f:one2} shows that the time evolution of the electrostatic potential exhibits a structure of alternating negative and positive polarity vortices along the poloidal (or $y$) direction, indicating that the poloidal average for the nonlinear term is weak. At the same time the vortices are not symmetric about the $x=0$ axis and these asymmetries result in the vorticity-like statistics.

\clearpage


\subsection{Two adjacent anti-parallel streams}
\label{s:two}

\begin{figure}[t]
\centerline{
\includegraphics[width=8cm, height=5.5cm]{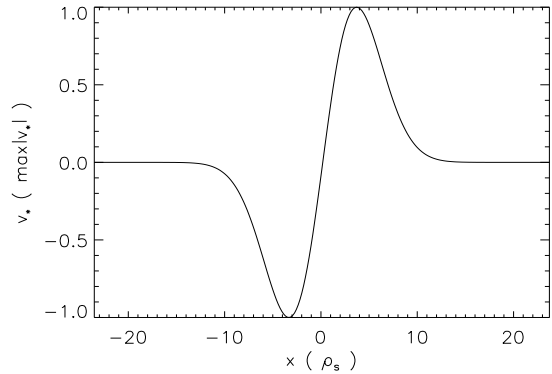}}
\caption{Constant $x$ profile of diamagnetic velocity $v_\star$ normalised 
to $\max|v_\star|$ for two anti-parallel streams. For slow flows the 
characteristic time is $(c_sL_n^{-1})^{-1}=3.8\times 10^{-5}$ s and  
for fast flows it is $(c_sL_n^{-1})^{-1}=10^{-5}$ s.
}
\label{f:slow1}
\end{figure}

The background density gradient (Figure \ref{f:slow1}) generates two streams 
flowing anti-parallel to each other with $\max|v_\star|$ centred at 
$x=\pm 3.5\rho_s$. The two streams flow adjacent to each other so that the 
edges of the streams are at $x=\pm 14\rho_s$ and $x=0$, where $v_\star$ 
becomes zero. Two cases with different $\max|v_\star|$ are studied: a 
slow-flowing and a fast-flowing system. 
For the slow-flowing streams the characteristic length and time give 
$\max|v_\star|=11.2\times 10^4$ cm s$^{-1}$. 
The fast-flowing streams have a $\max|v_\star|$ of 3.6 times that of 
the slow-flowing streams. Consequently the diamagnetic velocity gradient 
is larger for the fast-flowing streams. 

\begin{figure}[b]
\centerline{
\includegraphics[width=8.4cm, height=5.6cm]{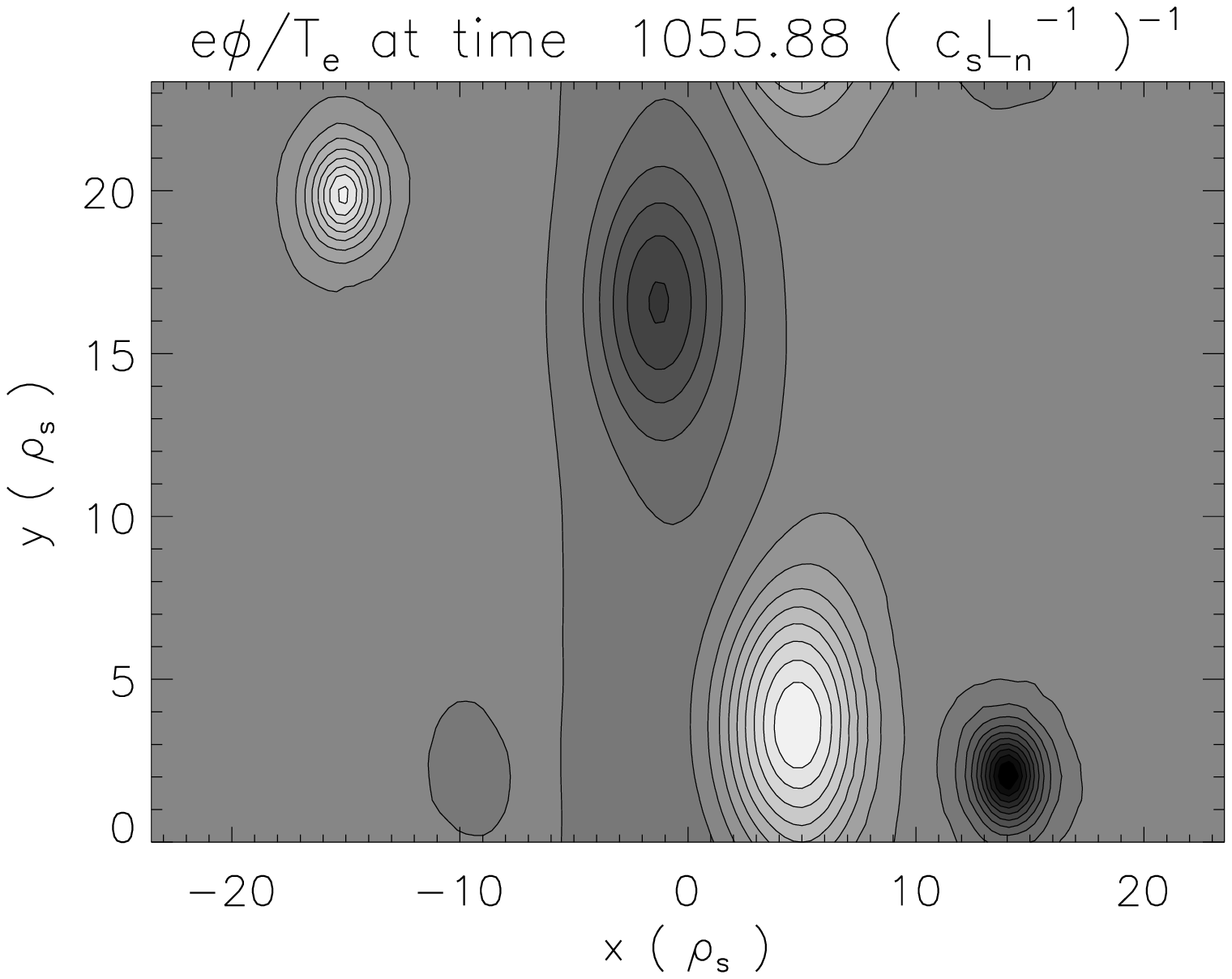}
\includegraphics[width=8.4cm, height=5.6cm]{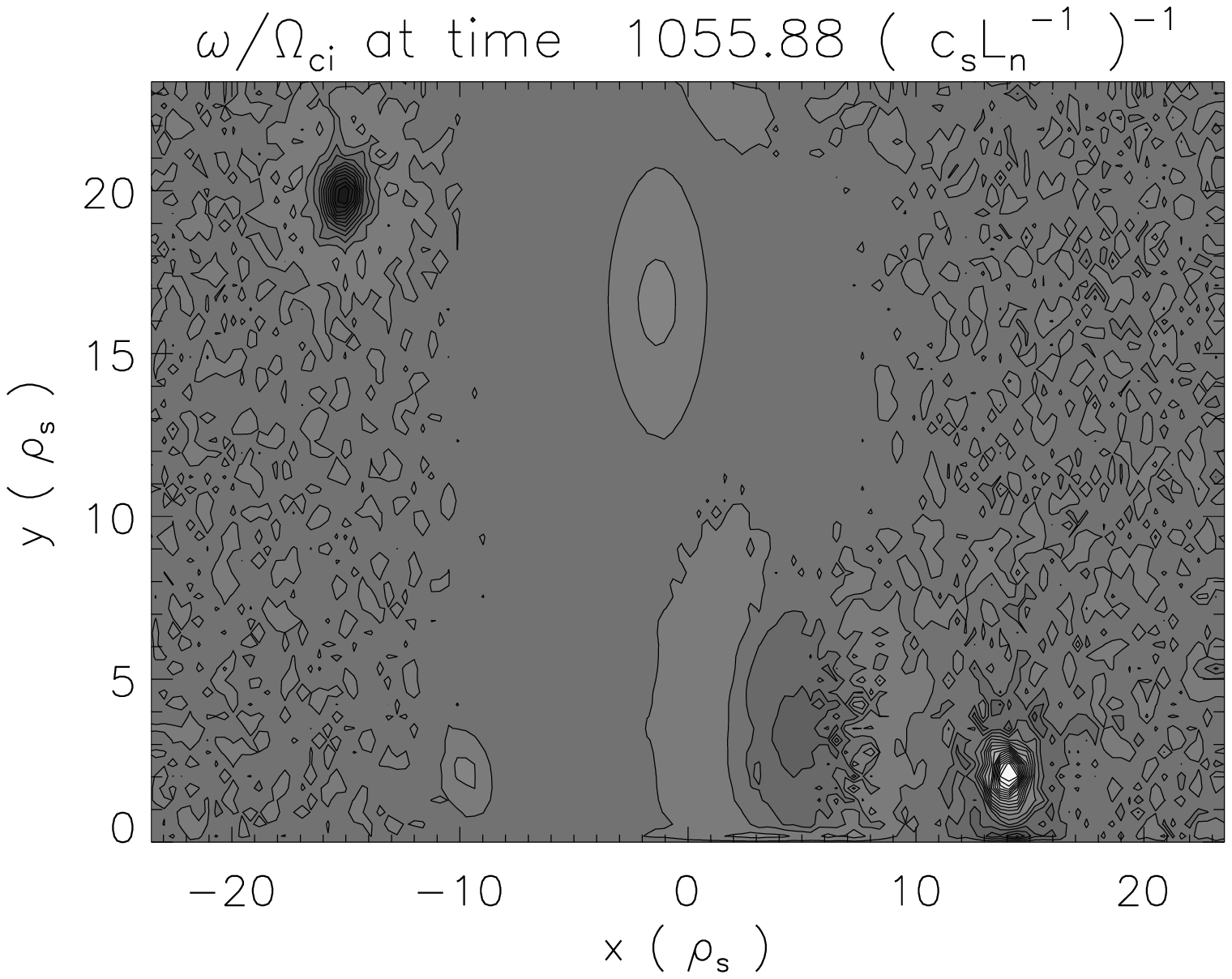} }
\caption{
The normalised potential $e\phi/T_e$ and normalised vorticity 
$\omega/\Omega_{ci}$ at 1055.88 characteristic time units for the 
slow-flowing streams. The minimum and maximum values of $e\phi/T_e$ are 
$-1.7\times 10^{-2}$ and $1.4\times 10^{-2}$.  
The extrema for $\omega/\Omega_{ci}$ are $-9.3\times 10^{-2}$ and 0.13.
Maximum is white and minimum black in both plots. 
The positive vortex at $x=5\rho_s$ moves in the negative $y$ direction at an 
approximate speed $3.4\max|v_\star|$, while the negative vortex at $x=0$ moves 
at approximate $2\max|v_\star|$ in the negative $y$ direction. It accelerates 
and decelerates every time the positive vortex passes it on its right hand 
side. The positive vortex at $x=-15\rho_s$ drifts slowly and randomly, while 
the negative vortex at $x=15\rho_s$ moves at speed $0.4\max|v_\star|$ in 
the positive $y$ direction. 
}
\label{f:slow2}
\end{figure}

\begin{figure}
\centerline{
\includegraphics[width=8.4cm, height=5.6cm]{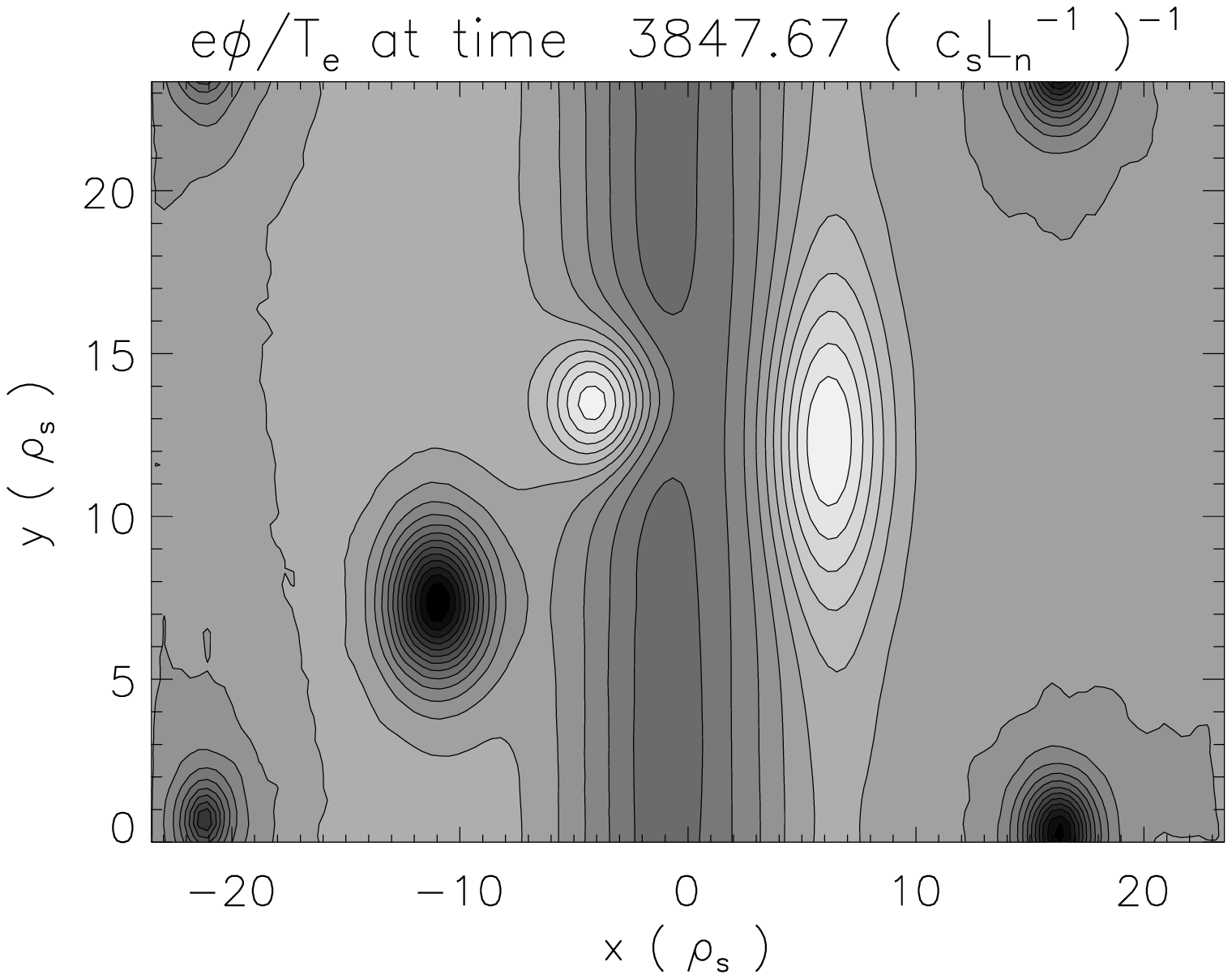}
\includegraphics[width=8.4cm, height=5.6cm]{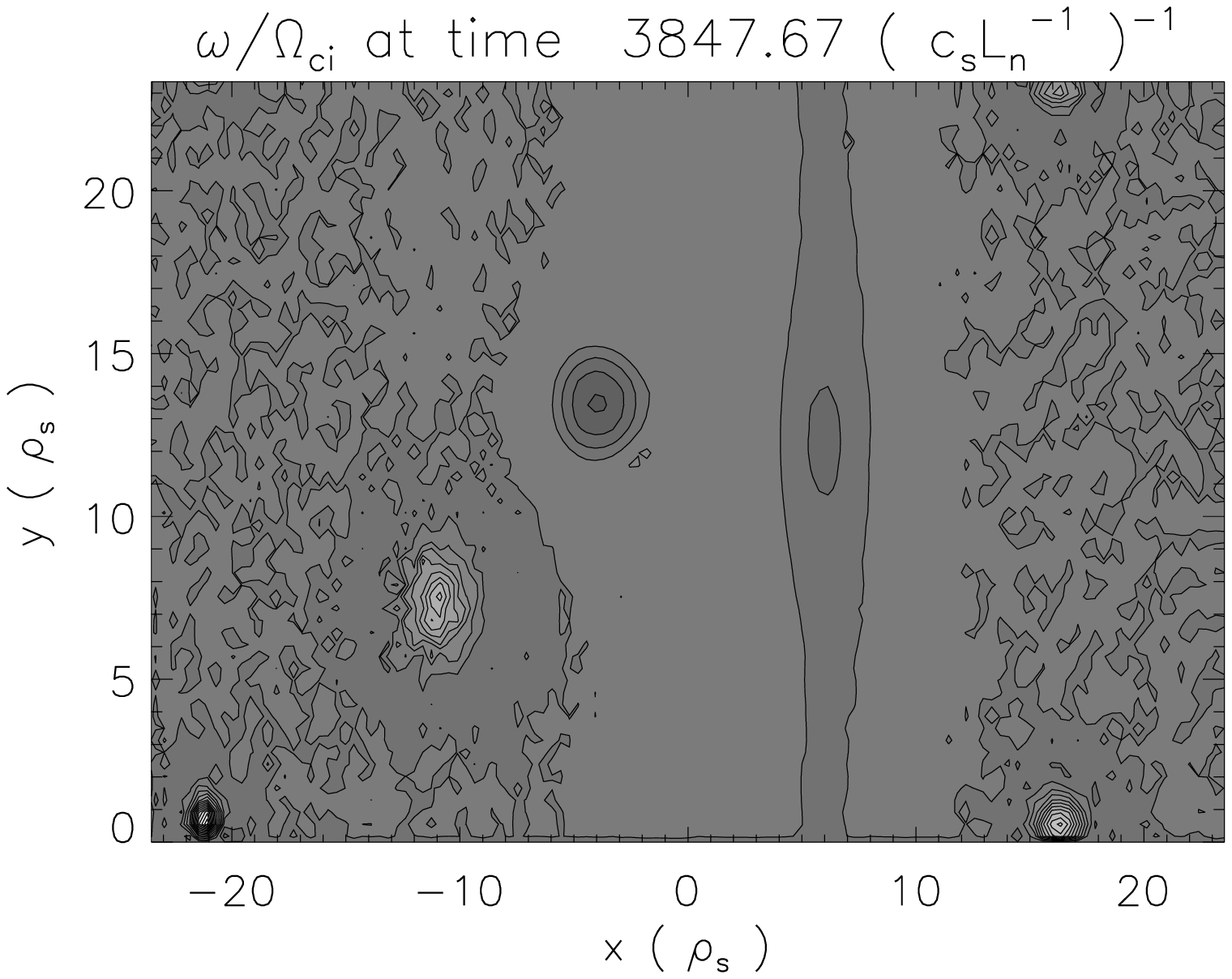} }
\caption{
The normalised potential $e\phi/T_e$ and normalised vorticity 
$\omega/\Omega_{ci}$ at 3847.67 characteristic time units for the 
fast-flowing streams. The minimum and maximum values of $e\phi/T_e$ are 
$-1.8\times 10^{-2}$ and $10^{-2}$, while the extrema for $\omega/\Omega_{ci}$ 
are $-3.9\times 10^{-2}$ and 0.18.
Maximum is white and minimum black in both plots. 
The negative vortex at $x=-11\rho_s$ moves at speed $0.9\max|v_\star|$ and the 
positive vortex at $x=-4\rho_s$ moves at $\max|v_\star|$, both in the positive 
$y$ direction. The positive vortex at $x=6\rho_s$ moves at speed 
$0.4\max|v_\star|$ in the negative $y$ direction while the negative vortex 
at $x=16\rho_s$ undergoes small and slow random movements.
}
\label{f:fast2}
\end{figure}

As in the case of one stream (Section \ref{s:one}), the initialisation 
creates many vortices that merge to form monopolar vortices moving in each 
of the streams. In the case of the slow-flowing streams, a smaller gradient 
in the diamagnetic velocity shear exists between the two streams, compared 
to the shear in the case of the fast-flowing streams. As a result, there are 
more interactions between the vortices in the anti-parallel streams in the case of the slow-flowing streams. 
This manifests itself in Figure \ref{f:slow2} by the fact that the two vortices situated inside the different streams move in the same direction, while the two vortices in Figure \ref{f:fast2} move in opposite directions. 

Positive vortices do not move in the direction of the stream where they are 
situated. Instead, they tend to move in the direction of the stream that is 
sampled by the edge of the vortex. This is clearly shown in Figures 
\ref{f:slow2} and \ref{f:fast2}. The negative vortices are dragged in the 
direction of their neighbouring positive vortices, shown by the vortex at 
$x=0$ in Figure \ref{f:slow2} and the one at $x=-11\rho_s$ in Figure 
\ref{f:fast2}. When the negative vortices are far enough from a positive 
vortex, their movement is determined by the flow direction sampled by 
their edge, as shown by the vortex at $x=15\rho_s$ in Figure \ref{f:slow2}. 
When vortices are far removed from the central flow streams and the 
vortices residing there, they exhibit small random movements, shown by 
the positive vortex at $x=-15\rho_s$ in Figure \ref{f:slow2} and the negative 
vortex at $x=16\rho_s$ in Figure \ref{f:fast2}. 

The generalised energy and enstrophy conservation follow the same pattern as 
for one stream (Section \ref{s:one}): after initialisation and large changes 
in energy and enstrophy the simulation produces a solution with less changes 
in these quantities. This is shown in Table \ref{t:slowfast} where the change 
measured from 40\% of the duration to the end of the time-line is 
approximately half the change measured from 20\% to the end.
The energy and enstrophy change throughout the run as the two adjacent 
streams influence each other. There is less interaction between the two 
streams when the diamagnetic velocity gradient between them is larger. 
This manifests in slightly better conservation in the faster streams 
(Table \ref{t:slowfast}).  

\begin{table}
\begin{tabular}{lrrcrr}
\hline
&
\multicolumn{2}{c}{$\bigtriangleup E/E_t=(E_t-E_1)/E_t$} & $\quad$ &
\multicolumn{2}{c}{$\bigtriangleup U/U_t=(U_t-U_1)/U_t$}\\
& $t=0.2$ & $t=0.4$ & & $t=0.2$ & $t=0.4$ \\ \hline
Slow-flowing streams & $0.50$ & $0.29$ & & $0.63$ & $0.37$ \\
Fast-flowing streams & $0.41$ & $0.26$ & & $0.52$ & $0.30$\\
\hline
\end{tabular}
\vspace{1cm}
\caption
{Conservation of the generalised energy $E$ and generalised enstrophy $U$ for 
two anti-parallel streams.
The time parameter $t$ is scaled such that $t=1$ is the end of each simulation.}
\label{t:slowfast}
\end{table}

\begin{figure}[h]
(a)
{\includegraphics[width=5.5cm, height=5.5cm]{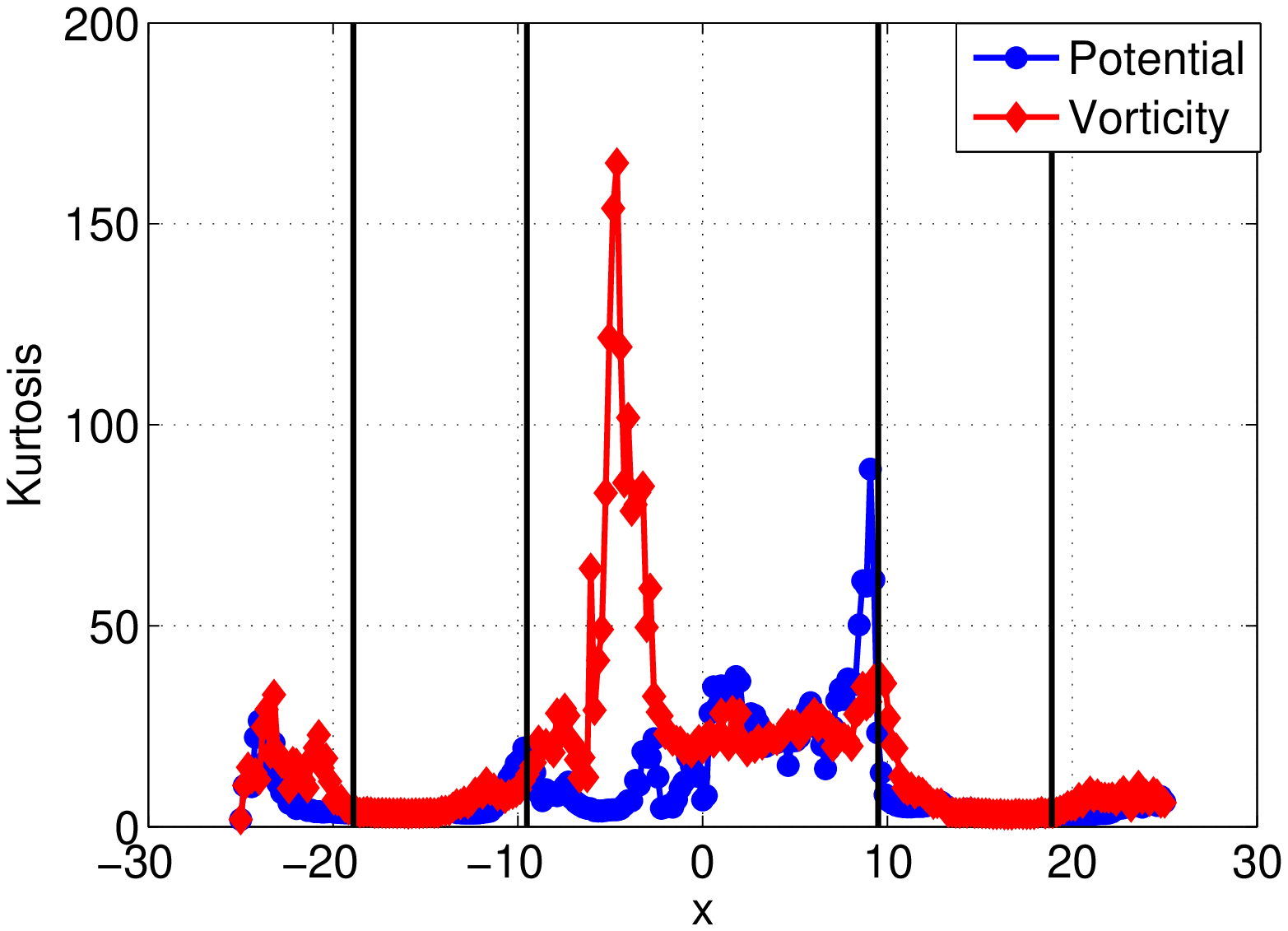}}
\hspace{2cm}
(b)
{\includegraphics[width=5.5cm, height=5.5cm]{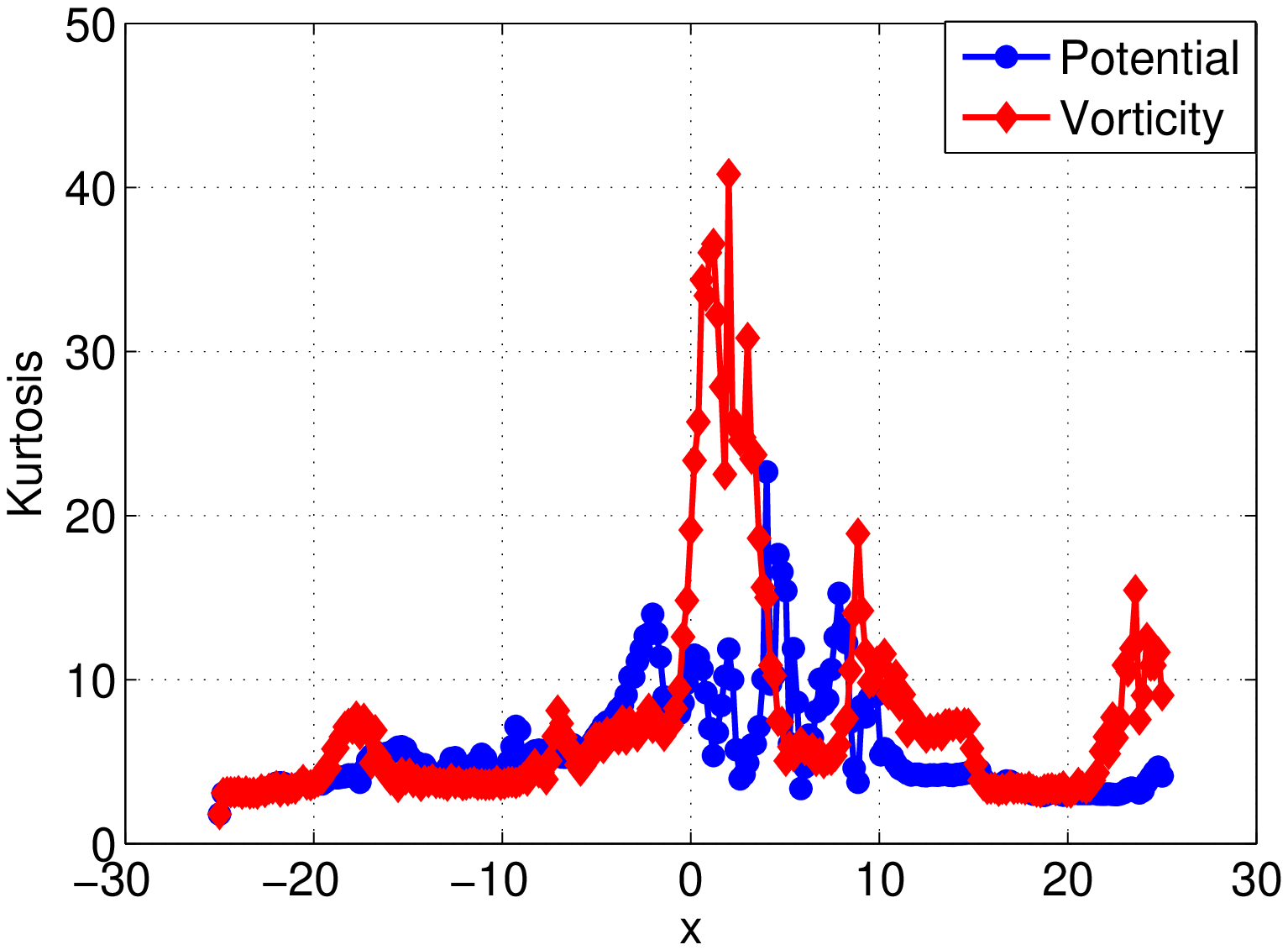}}
\caption{
The kurtosis of the potential and vorticity of the ARIMA modeled time traces along the $x$ direction are shown for (a) the slow-flowing and 
(b) the fast-flowing anti-parallel cases.}

\label{f:Kurt_slowfast}
\end{figure}

As a comparison to the statistical analysis of the one stream case, we will consider the time evolution of the stochastic part of the electrostatic potential and vorticity for both slow-flowing and fast-flowing anti-parallel streams. In Figure \ref{f:Kurt_slowfast}(a), the kurtosis of the potential and vorticity time traces are displayed corresponding to the simulation of the slow flow presented in Figure \ref{f:slow2}. 
  
We find similarly good correspondence in kurtosis profiles between the potential and vorticity as was found in the case of one stream (Section 
\ref{s:one}) for the stochastic residual part whereas for the original time traces no such correspondence could be found. We have omitted the figures of the kurtosis of original time series due to space limitations here and in the rest of the paper, since these do not provide any additional useful information.

\begin{figure}[ht]
{\vspace{2mm}
\includegraphics[width=8.1cm, height=8.7cm]{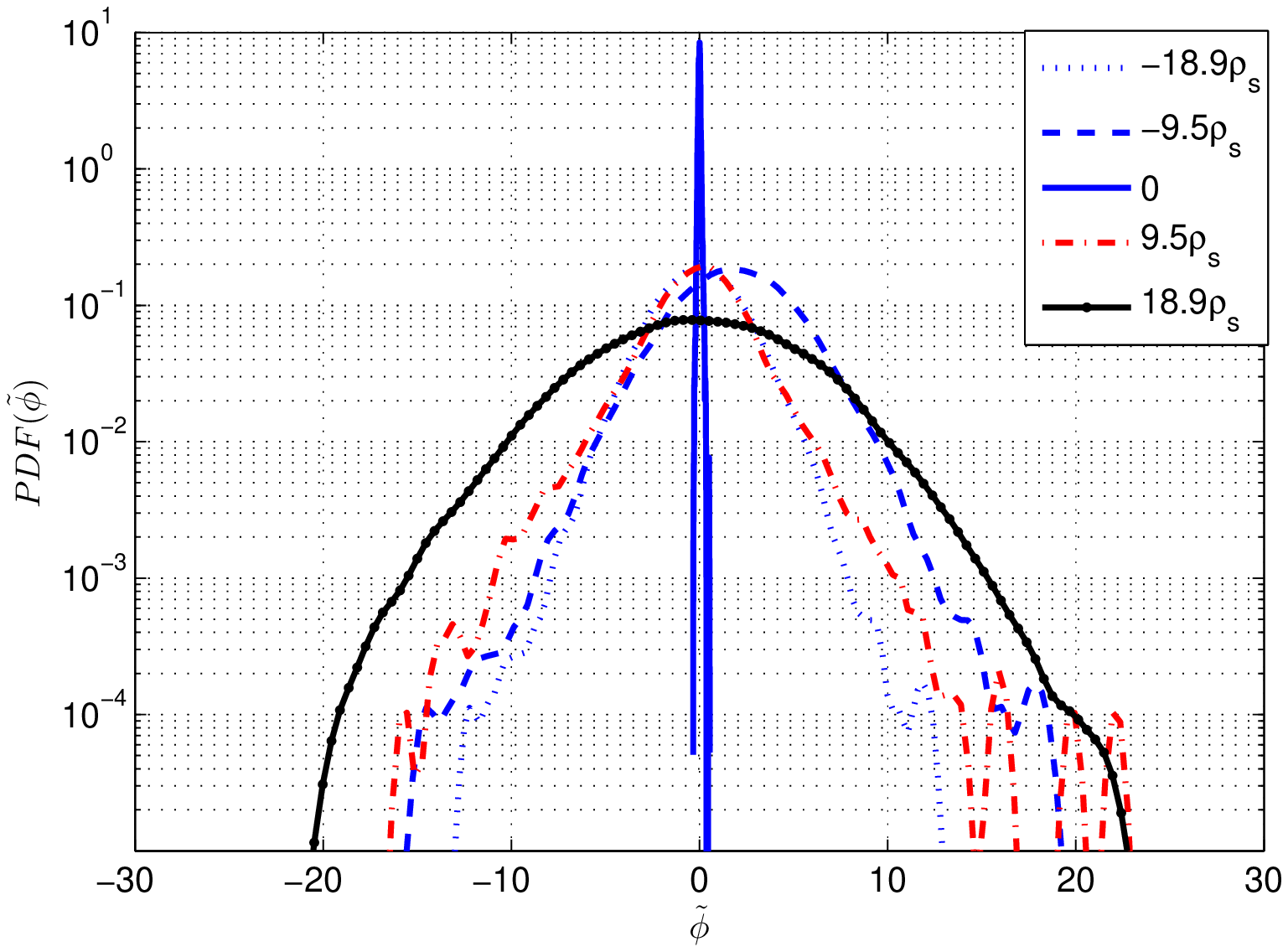}}
{\includegraphics[width=8.1cm, height=8.7cm]{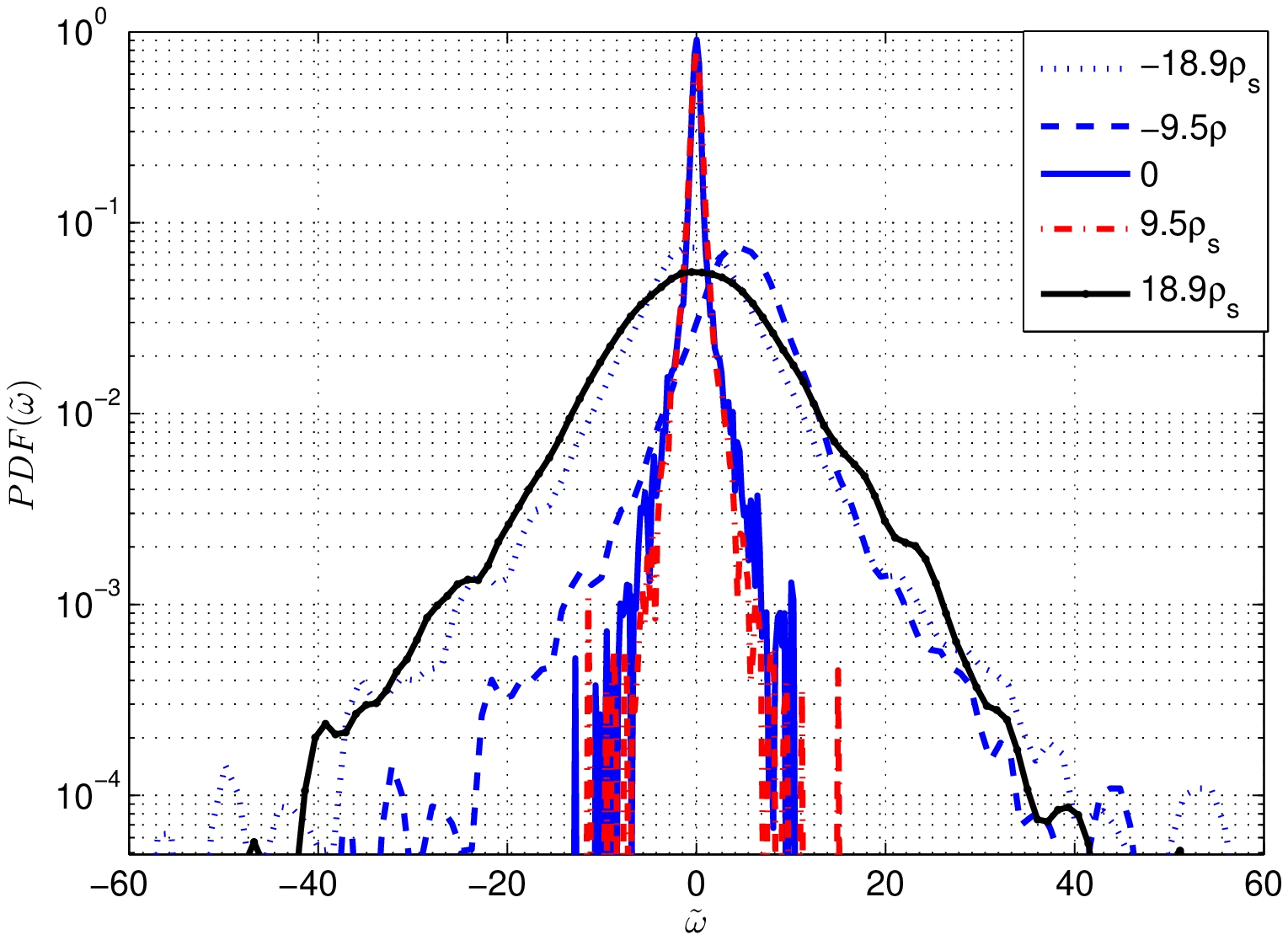}}
\caption{
The PDFs of the potential (graph on left) and vorticity (graph on right) time traces along the $x$ direction in the slow-flowing anti-parallel case. The various $x$ positions sampled are indicated by black lines in Figure \ref{f:Kurt_slowfast}(a).
}
\label{f:PDFs_slow}
\end{figure}

\begin{figure}[ht]
{\vspace{2mm}
\includegraphics[width=8.1cm, height=8.7cm]{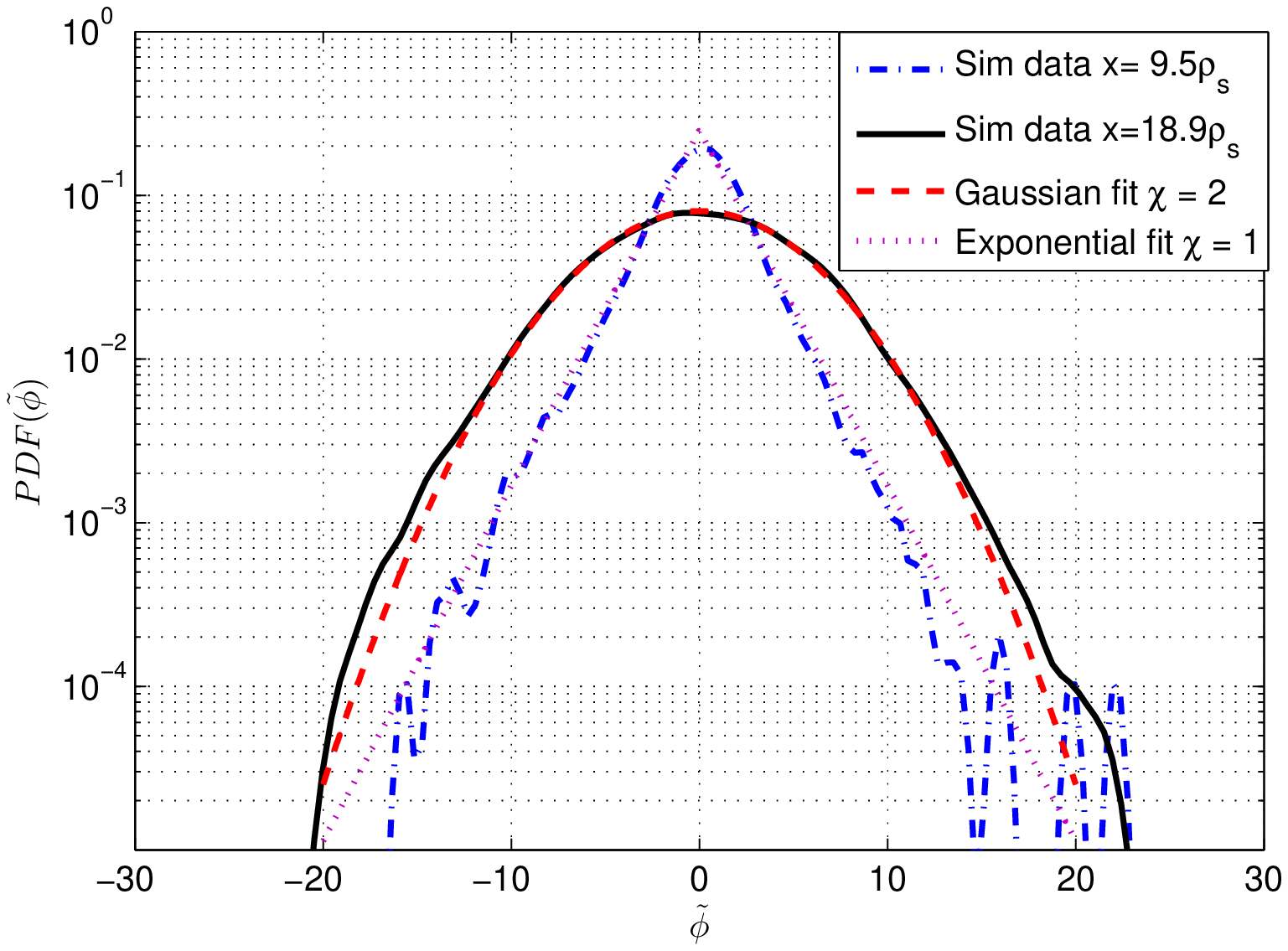}
}
\caption{
Fits to the two salient types of PDFs, the Gaussian and the Laplacian (exponential) distributions.
}
\label{f:PDFs_slow_fit}
\end{figure}

In Figure \ref{f:PDFs_slow} and \ref{f:PDFs_slow_fit}, the numerical PDFs of the stochastic part $\phi_{res}$ are shown for the CHM simulations of Figure \ref{f:slow2}. We find that at the edge of the stream the PDFs are close to Gaussian whereas at the stream centre other nonlinear features can be found. Here we will utilize Eq. (\ref{pq2}) in the following cases: Laplacian distribution denotes the analytical model for $\chi = 1.0$, whereas a Gaussian PDF is represented by $\chi = 2.0$. The appearance of a Laplacian distribution at the stream (Figure \ref{f:PDFs_slow_fit}) is suggestive of a vorticity conserving nonlinear system \cite{falcovich2011, anderson4}, whereas a Gaussian distribution is likely for a weakly nonlinear system \cite{anderson2} or dynamics impeded by a strong zonal flow \cite{anderson3}.  

In Figure \ref{f:Kurt_slowfast}(b) the kurtosis of the time series generated by the fast-flowing anti-parallel streams corresponding to the simulation presented in Figure \ref{f:fast2}. The ARIMA modeled time trace of potential and vorticity follow each other reasonably well compared to the original time traces. Also here the skewness is small and the PDFs are Gaussian or exponential.

\clearpage


\subsection{Two streams with space between them}
\label{s:space}

\begin{figure}[t]
\centerline{
\includegraphics[width=8.4cm, height=5.6cm]{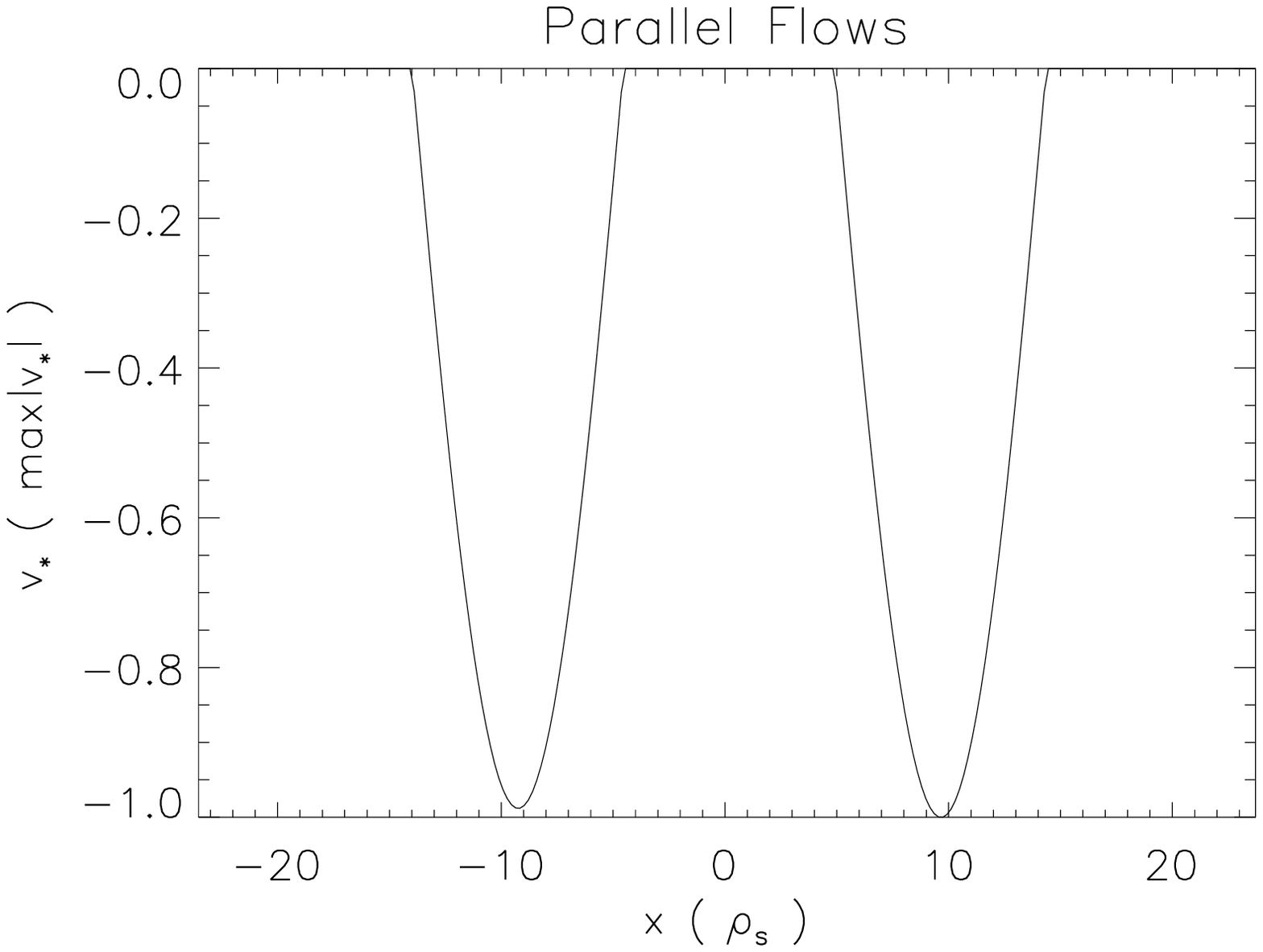}
\includegraphics[width=8.4cm, height=5.6cm]{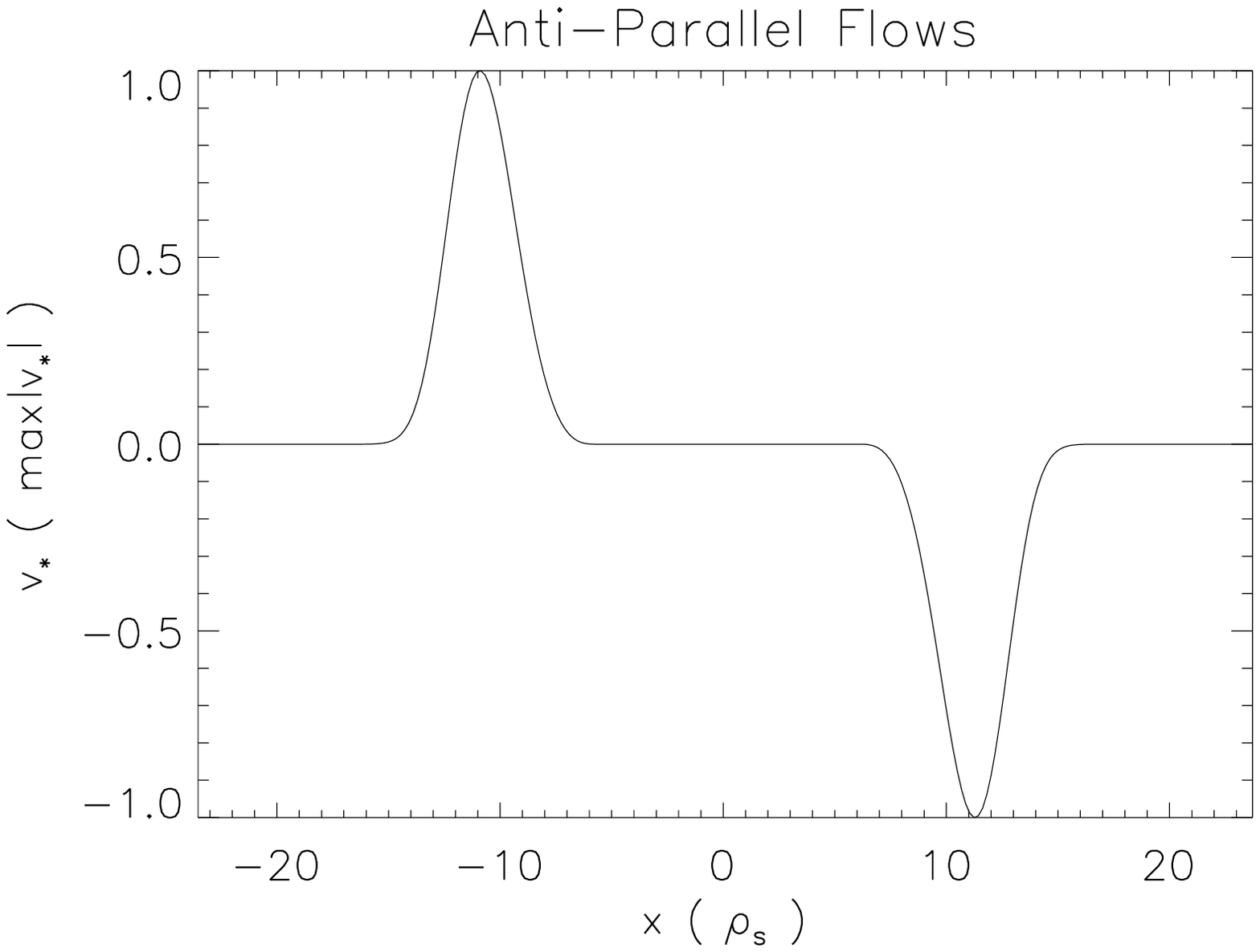} }
\caption{Constant $x$ profiles of diamagnetic velocity $v_\star$ normalised 
to $\max|v_\star|$ for two streams with a space between them. 
The characteristic time is $(c_sL_n^{-1})^{-1}=8.9\times 10^{-6}$ s 
for both parallel and anti-parallel streams.}
\label{f:para1}
\end{figure}

Figure \ref{f:para1} shows the background density gradient generating 
the two cases studied here: two streams flowing parallel to each other and 
two streams flowing anti-parallel to each other. For both cases there exists a 
space between the streams. 
The parallel streams are located at positions $x\in[-14,-4]$ and $x\in[4,14]$ 
with $\max|v_\star|$ at $x=\pm 9\rho_s$ and $v_\star=0$ outside these intervals.
The anti-parallel streams are located at positions $x\in[-16,-6]$ and 
$x\in[6,16]$ with $\max|v_\star|$ at $x=\pm 11\rho_s$ and $v_\star=0$ outside 
these intervals. The characteristic length and time are the same in both 
cases, giving $\max|v_\star|=4.7\times 10^4$ cm s$^{-1}$.

\begin{figure}[t]
\centerline{
\includegraphics[width=8.4cm, height=5.6cm]{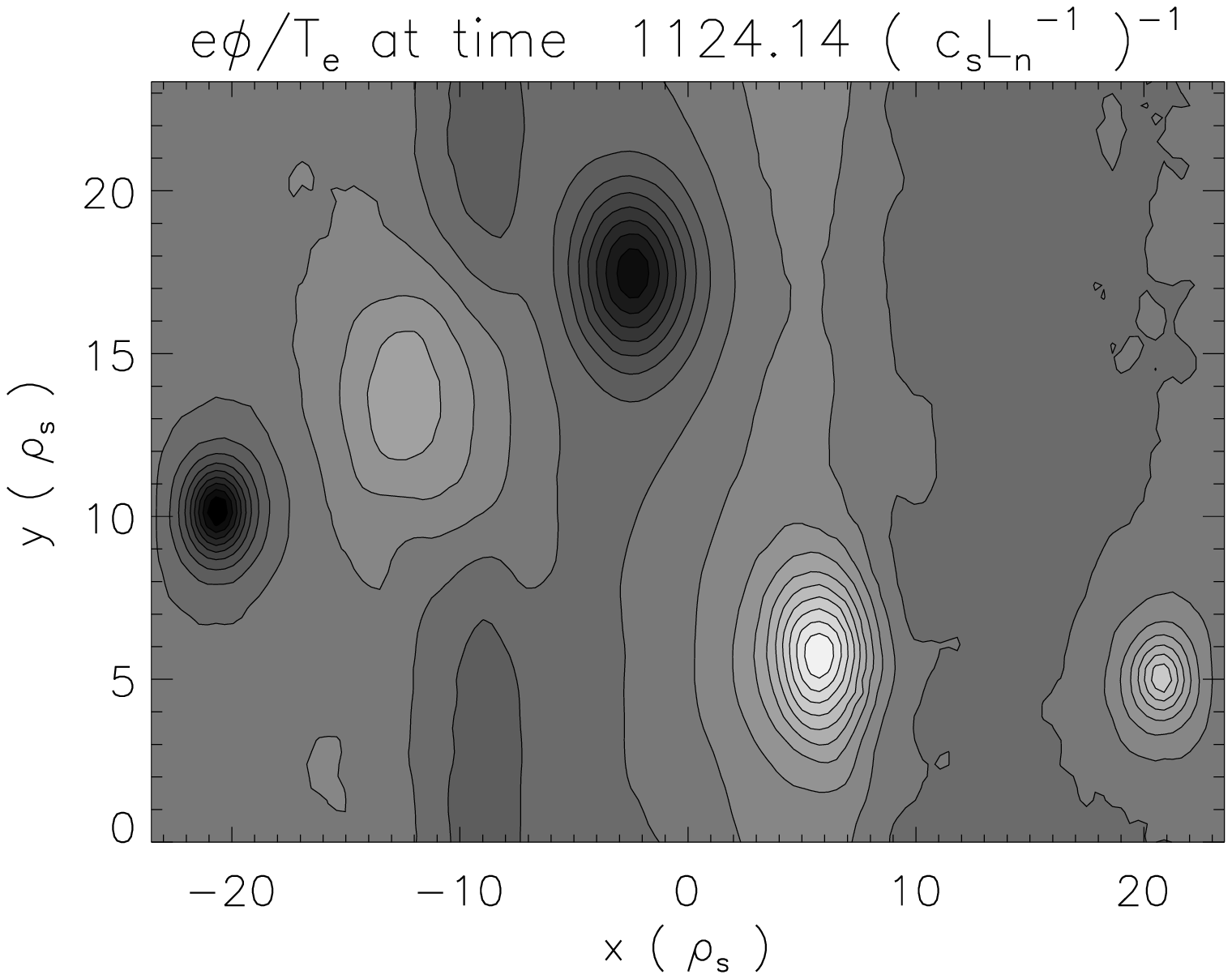}
\includegraphics[width=8.4cm, height=5.6cm]{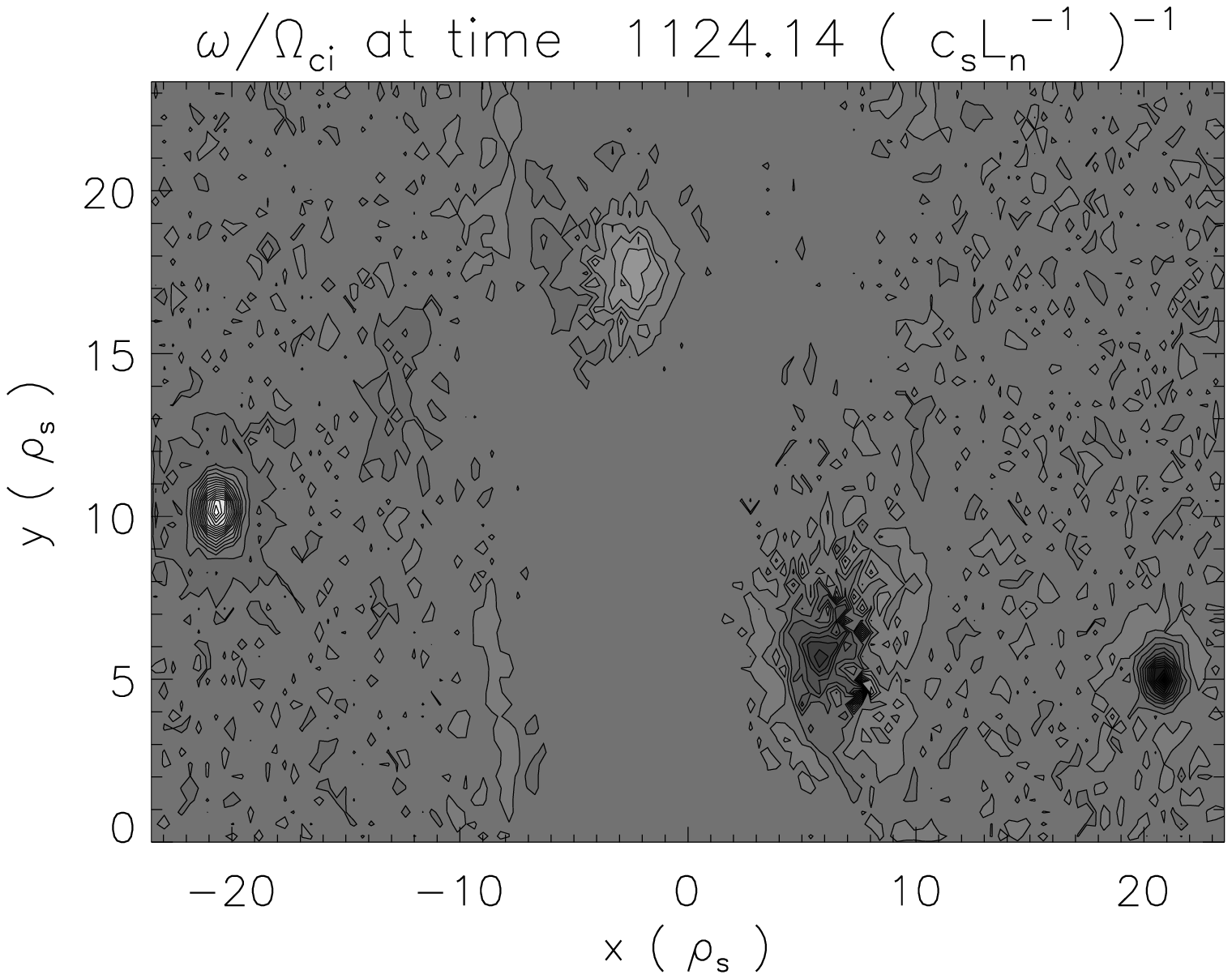} }
\caption{
The normalised potential $e\phi/T_e$ and normalised vorticity 
$\omega/\Omega_{ci}$ at 1124.14 characteristic time units for two parallel 
streams. The minimum and maximum values of $e\phi/T_e$ are 
$-2.3\times 10^{-2}$ and $2.4\times 10^{-2}$, while the extrema for 
$\omega/\Omega_{ci}$ are 0.18 and 0.2.
Maximum is white and minimum black in both plots. 
The negative vortex at $x=-3\rho_s$ moves at speed $0.7\max|v_\star|$ and the 
positive vortex at $x=6\rho_s$ moves at $1.2\max|v_\star|$, both in the negative 
$y$ direction. 
}
\label{f:para2}
\end{figure}

Initial pairs of bipolar vortices merge to form monopolar vortices. 
In the case of two parallel flowing streams, all the vortices move in the 
flow direction of the two streams (Figure \ref{f:para2}). 
These vortices exhibit the same behaviour as those in Section \ref{s:two}, 
namely they move in the direction of the flow sampled by their edges. 
Vortices outside the two streams are not influenced by the 
flow direction of the streams, e.g., the vortices at $x=\pm20\rho_s$ in 
Figure \ref{f:para2}.  

\begin{figure}[t]
\centerline{
\includegraphics[width=8.4cm, height=5.6cm]{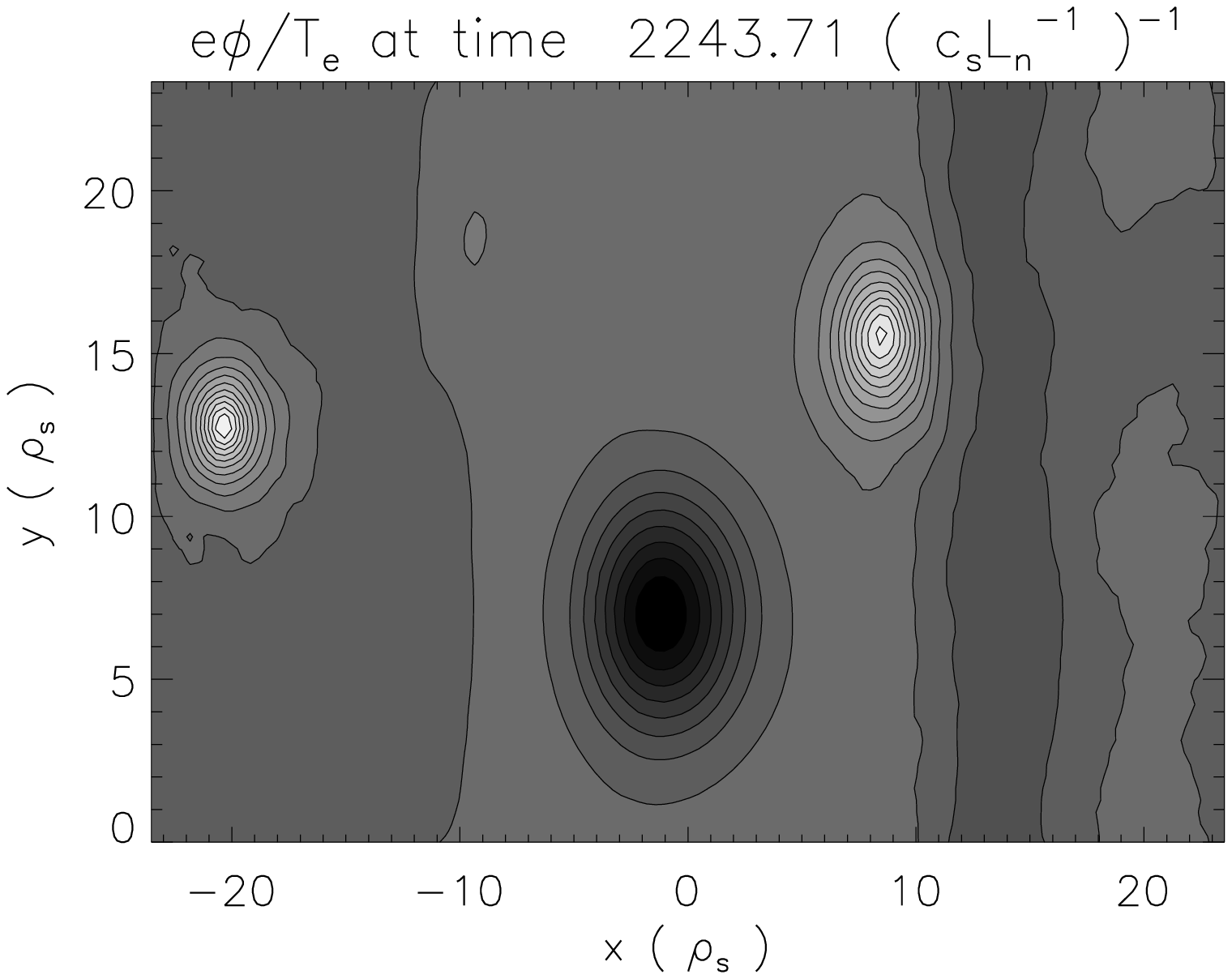}
\includegraphics[width=8.4cm, height=5.6cm]{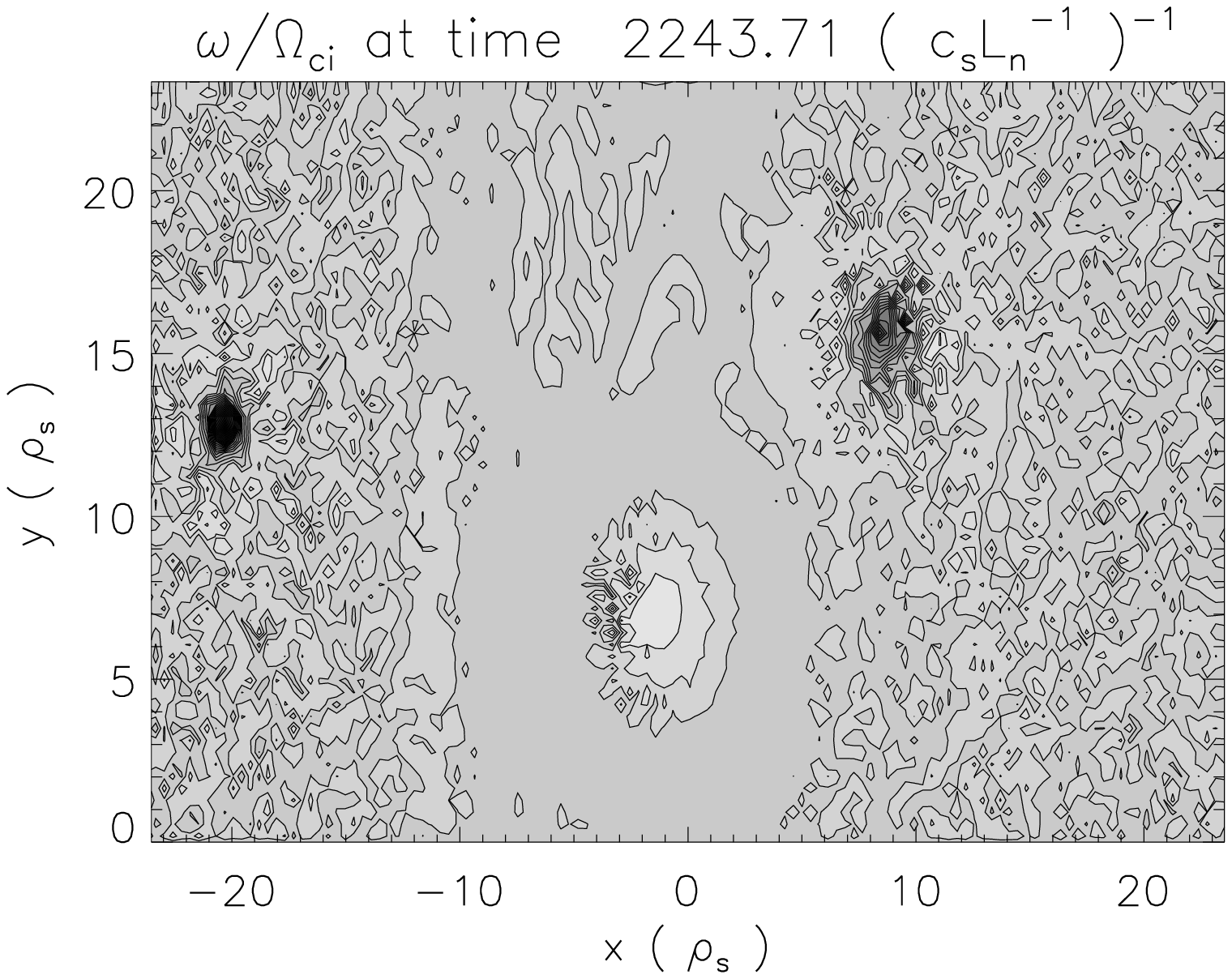} }
\caption{
The normalised potential $e\phi/T_e$ and normalised vorticity 
$\omega/\Omega_{ci}$ at 2273.71 characteristic time units for anti-parallel 
streams. The minimum and maximum values of $e\phi/T_e$ are 
$-1.8\times 10^{-2}$ and $2.5\times 10^{-2}$, while the extrema for 
$\omega/\Omega_{ci}$ are -0.25 and $5.2\times 10^{-2}$.
Maximum is white and minimum black in both plots. 
The negative vortex at $x=-\rho_s$ moves at speed $1.4\max|v_\star|$ in the 
negative $y$ direction and the positive vortex at $x=8\rho_s$ moves at speed 
$0.6\max|v_\star|$ in the positive $y$ direction. 
The positive vortex at $x=-21\rho_s$ hardly moves at all.
}
\label{f:anti2}
\end{figure}

In the case of the two anti-parallel flowing streams, monopolar vortices 
form inside the streams from bipolar vortices after initialisation. These 
vortices move in the opposite direction of the stream in which they reside. 
Monopolar vortices forming between the two anti-parallel streams at positions 
$x>0$ move in the flow direction their edges sample, i.e., the negative 
$y$ direction. At some stage during the simulation these vortices migrate in 
the negative $x$ direction, moving through the stream situated at $x=-10\rho_s$ 
and in the process destroying the vortices in that stream. The result is 
shown in Figure \ref{f:anti2} where one vortex, situated at $x=-\rho_s$, moves 
between the streams in the negative $y$ direction while another moves in the 
opposite direction of the stream flow at $x=8\rho_s$.  

Table \ref{t:antipara} shows that the generalised energy conservation is good 
for the two streams flowing in the same direction, but bad for the 
anti-parallel streams. This shows that the amplitude of the fluctuations 
decrease more in the latter case. There is no discernible pattern in the 
conservation of the generalised enstrophy.

\begin{table}
\begin{tabular}{lrrcrr}
\hline
&
\multicolumn{2}{c}{$\bigtriangleup E/E_t=(E_t-E_1)/E_t$} & $\quad$ &
\multicolumn{2}{c}{$\bigtriangleup U/U_t=(U_t-U_1)/U_t$}\\
& $t=0.2$ & $t=0.4$ & & $t=0.2$ & $t=0.4$ \\ \hline
Parallel streams      & $0.03$ & $-0.003$ & & $0.007$ & $-0.21$ \\
Anti-parallel streams & $0.32$ & $0.160$ & & $0.274$ & $0.02$\\
\hline
\end{tabular}
\vspace{1cm}
\caption
{Conservation of the generalised energy $E$ and generalised enstrophy $U$ for 
two streams with a space between them.
The time parameter $t$ is scaled such that $t=1$ is the end of each simulation.}
\label{t:antipara}
\end{table}

\begin{figure}[h]
(a)
{\includegraphics[width=5.5cm, height=5.5cm]{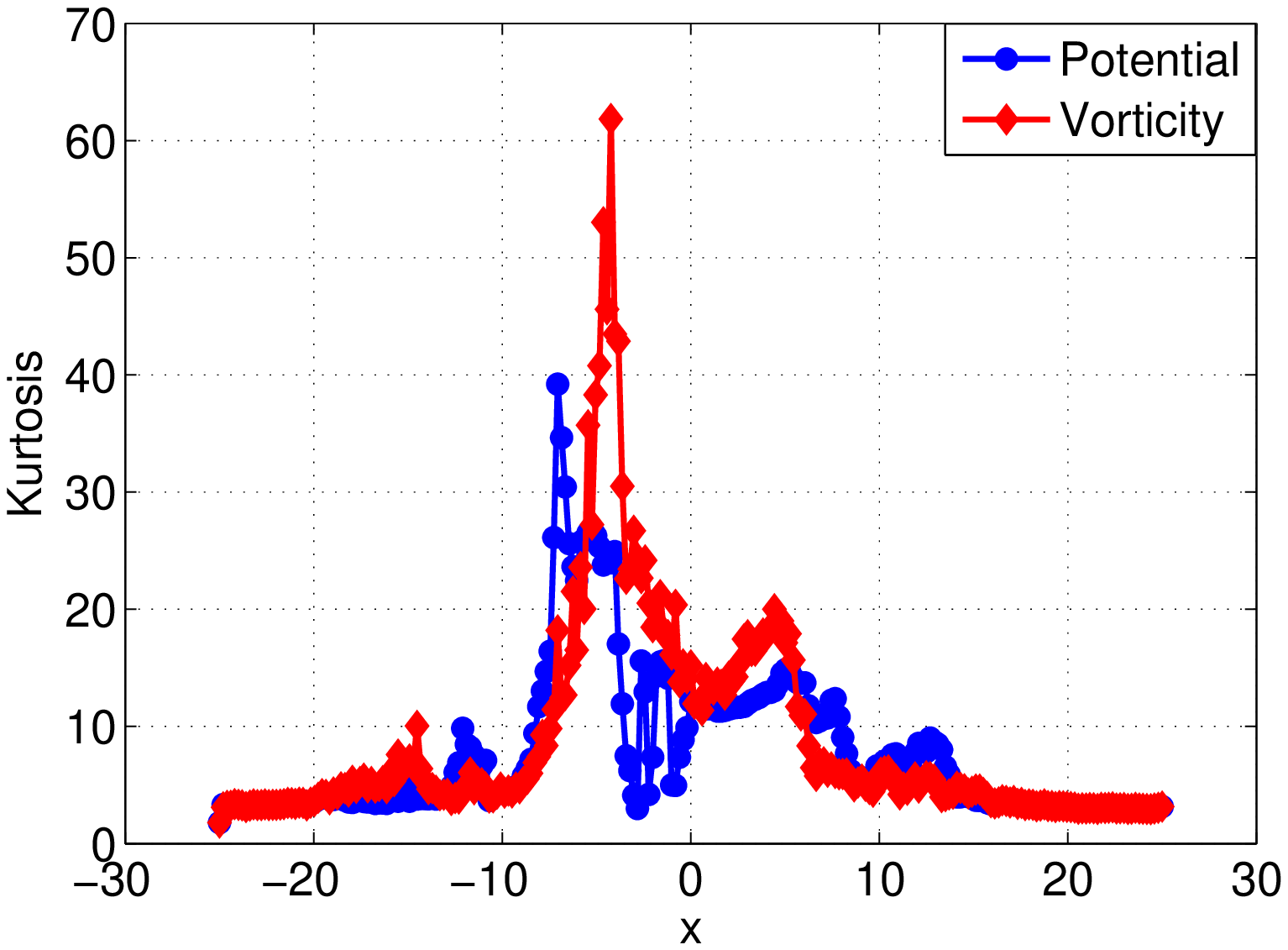}}
\hspace{2cm}
(b)
{\includegraphics[width=5.5cm, height=5.5cm]{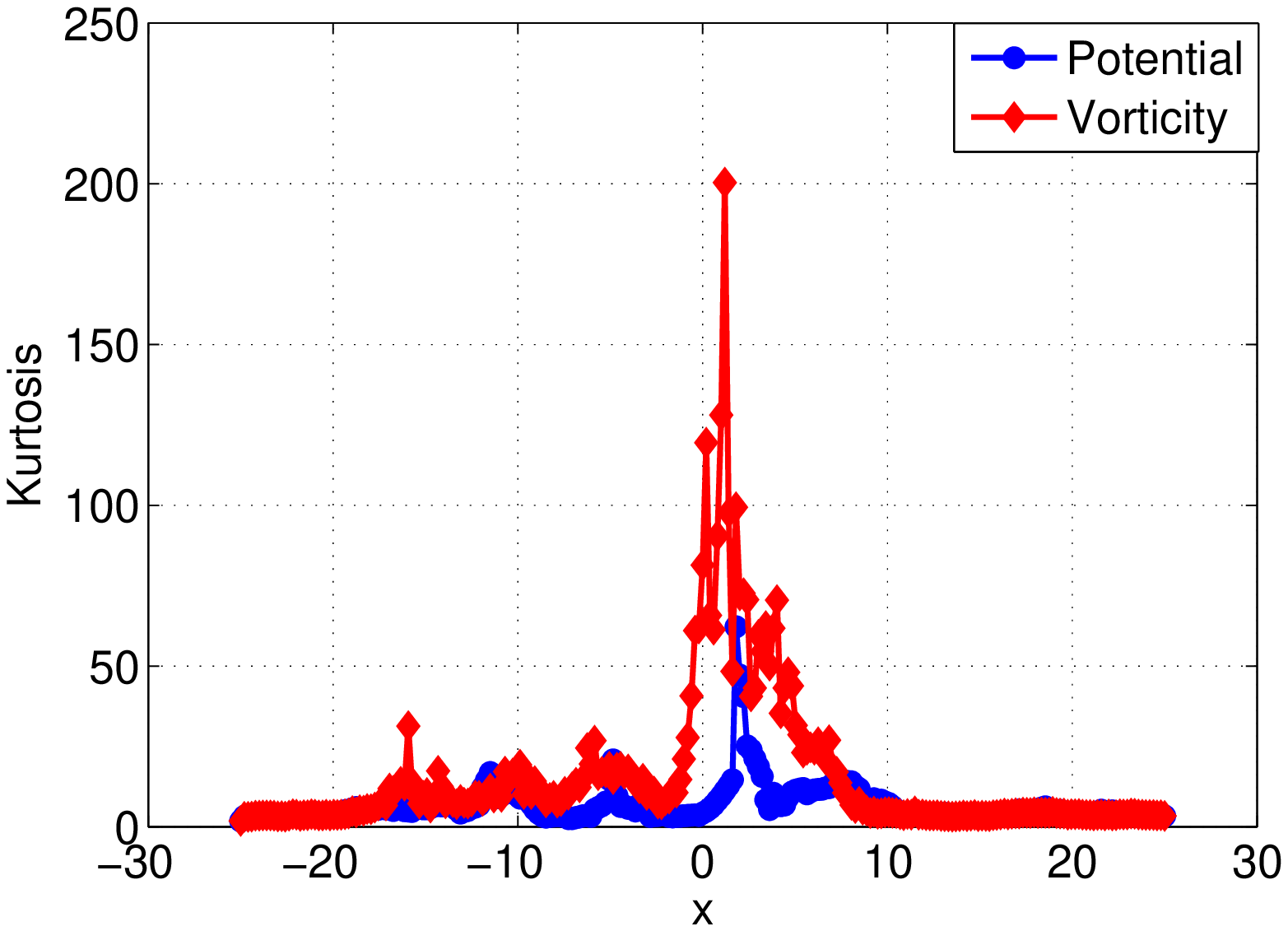}}
\caption{
The kurtosis of the potential and vorticity of the ARIMA modeled time traces along the $x$ direction are shown in the (a) parallel and (b) anti-parallel cases.
}
\label{f:Kurt_apara}
\end{figure}

Figure \ref{f:Kurt_apara} presents the kurtosis 
generated with two imposed parallel or anti-parallel flows with a spacing 
between them. 
The statistics are obtained from the time evolution of the electrostatic 
potential in Figures \ref{f:para2} and \ref{f:anti2} for the parallel and 
anti-parallel flows respectively. 
Note that the kurtosis of time traces of potential and vorticity follow 
each other reasonably well compared to the original time traces.

\clearpage 


\section{Discussion and Summary}
\label{sec:sum}

In this paper simulations of unforced Charney-Hasegawa-Mima (CHM) flows 
have been performed 
where we have imposed constant background density profiles on the CHM 
equation. In this manner we generated a single zonal flow, anti-parallel   
flows with slow and fast moving plasma adjacent to each other, as well as 
parallel and anti-parallel flows with a space between the flow streams. 

In the single flow stream a chain of monopolar vortices form along the 
stream with their widths determined by the stream width and with 
polarities opposite from their neighbours. The vortex chain moves in 
the flow direction. 
When two anti-parallel zonal flows are placed adjacent to each other, 
monopolar vortices form in each stream but they are affected by the 
neighbouring stream and vortices in their close proximity. 
The smaller the flow velocity gradients are, the more interaction 
occurs. Vortices move in the direction their edges sample. 
In addition, negative polarity vortices in the presence of positive 
vortices are dragged along by the positive vortex. 
When two streams are placed a distance from each other, 
monopolar vortices form inside as well as between the streams. 
All vortices move in the flow direction when the streams flow in 
parallel. In the case of anti-parallel flow the vortex movement is 
in the opposite direction of the residing flow velocities. Vortex 
movement between the two flows are dictated by the flow direction 
the vortex edge samples.  

After initialisation bipolar vortices form that merge and destroy 
each other until only monopolar vortices survive, which leads to 
enstrophy changes during the initial phase. The amplitudes 
of the vortices also change to a stable value during this phase, 
hence an initial change in energy. Once the initial phase is over, 
the energy and enstrophy conservation is good for the one 
stream. In the case of the two stream scenarios the energy and 
enstrophy are conserved to a lesser extent in some numerical runs 
and to a larger extent in others. 
This is because vortices keep on being created and destroyed as 
they interact in a random manner with each other and the sheared flow.

We have sampled time series at points in the 
radial direction (poloidally averaged) of the electrostatic potential and 
corresponding vorticity generated by the simulations. Our aim is to evaluate 
the intermittent characteristics of the time series by using a standard 
Box-Jenkins modeling. This mathematical procedure effectively removes 
deterministic autocorrelations from the time series, 
allowing for the statistical interpretation of the stochastic residual part. 
The numerically generated time traces are compared with predictions from a 
nonperturbative theory, the so-called instanton method for computing 
probability density functions (PDFs) in turbulence. More specifically the 
numerically generated time traces are analysed using the ARIMA model and 
fitted with analytical models accordingly.
In the simulations presented here we find that an ARIMA(3,1,0) model 
presents an adequate description of the stochastic process.

The time series of the ARIMA modeled stochastic residual of cases described in section \ref{sec:results} (A,B,C)
of the potential and vorticity exhibit in general a uni-modal PDF with Gaussian features or a 
PDF with exponential tails. In summary, the PDF $\sim \exp\big(- const \ |\phi|^{\chi}\big)$ with $\chi=2.0$ or 
exponential statistics are found with ($\chi = 1.0$). The different configurations are
represented by an imposed  slow and a fast zonal flow as well as parallel and anti-parallel flows.
The rationale for using the ARIMA model is to uncover the stochastic process hidden in the numerically generated time trace. The ARIMA process is specifically designed to identify correlations in a time trace by utilizing a differencing procedure and thus provides us with an efficient model how to subtract the correlations in time from the full signal. The objective of this work is to identify the particular stochastic process that is generating the time trace. The one important restriction of this procedure is that the stochastic process has to be stationary with respect to the mean and variance. For an arbitrary stochastic process, there exist formal tests to check whether stationarity holds. In this particular example of CHM zonal flows with a constant background density profile generating the process, this is fulfilled except for a short interval at the start of the simulation.

One non-trivial aspect of the statistics of the electrostatic potential in the 
CHM system is conservation of energy and enstrophy that may lead to a Gibbsian equilibrium distribution. However with the imposed shear flow through the varying density gradient, non-vanishing triad interactions redistribute energy between the modes, contributing to a situation where the PDFs deviate from the Gaussian form at some radial locations. 
For a general discussion see Krommes\cite{krommes_kinetic}. The cascade processes or redistribution energy and enstrophy can also be directly seen in the simulations as deviations in the conserved quantities.  

To this end, in general we find an emergent universal exponential scaling 
of the distribution functions (Laplace distribution with $\chi = 1.0$) that 
accurately describes the statistics of the time series of the electric 
potential and vorticity. Analysing the profiles along the $x$ coordinate of 
the kurtosis of the potential and vorticity time series we find
striking similarity suggestive of the relation in Eq. (\ref{modrel}). 
Exceptions to the named distributions occur where strong nonlinear 
interactions are present in the dynamics where the PDFs are sub-exponential 
with $\chi < 1.0$ and high values of the normalized fourth moment 
(kurtosis) is found.



\end{document}